\documentclass{IEEEtran}

\usepackage{graphicx,epsf}
\usepackage{bbm,color}
\usepackage{amsmath,amssymb,amsfonts}
\usepackage{multirow}
\usepackage{bm}
\usepackage{url}
\usepackage[pdftex,colorlinks=true,urlcolor=magenta,citecolor=magenta,linkcolor=magenta]{hyperref}

\usepackage[numbers]{natbib}

\usepackage{graphicx}
\usepackage{subfig}

\newcommand{\E}{\mathbb E}
\newcommand{\real}{{\mathbb R}}

\newcommand{\N}{\mathcal{N}}

\newcommand{\kb}{{k}}

\newcommand{\var}{\makebox{ Var }}

\def\threed{\widetilde\lambda_{j,k}} 			

\def\r{\color{red}}        			

\newcommand{\miniskip}{\vspace{1mm}}

\graphicspath{
{Figures/}
{/Users/pabry/TEXTES/WORK_ON_LINE/CNRSJSPS/FontugneMF/MAWI_MF/Figures3Day/fig/estimates/}
{/Users/pabry/TEXTES/WORK_ON_LINE/CNRSJSPS/FontugneMF/MAWI_MF/Figures3Day/fig/each_day_LD/}
{/Users/pabry/TEXTES/WORK_ON_LINE/CNRSJSPS/FontugneMF/MAWI_MF/Figures3Day/fig/all_med_LD/}
{/Users/pabry/TEXTES/WORK_ON_LINE/CNRSJSPS/FontugneMF/MAWI_MF/Figures15min/fig/misc/}
{/Users/pabry/TEXTES/WORK_ON_LINE/CNRSJSPS/FontugneMF/MAWI_MF/Figures15min/fig/estimates/}
{/Users/pabry/TEXTES/WORK_ON_LINE/CNRSJSPS/FontugneMF/MAWI_MF/Figures15min/fig/each_day_LD/}
{/Users/pabry/TEXTES/WORK_ON_LINE/CNRSJSPS/FontugneMF/MAWI_MF/Figures15min/fig/all_med_LD/}
{/Users/pabry/TEXTES/PUBLIS/COURANT/15Publis/15TrafficMF/Figures/}
{/Users/pabry/TEXTES/PUBLIS/COURANT/15Publis/15ICASSP/15ICASSP_internet/ICASSP2015_MFI/}
}

\renewenvironment{IEEEbiography}[1]
  {\IEEEbiographynophoto{#1}}
  {\endIEEEbiographynophoto}

\begin{document}

\title{Scaling in Internet Traffic:\\ a 14 year and 3 day longitudinal study,\\ with multiscale analyses and random projections}

\author{Romain Fontugne$^{(1,4)}$, 
Patrice Abry$^{(2,1)}$,~\IEEEmembership{IEEE Fellow}, 
Kensuke Fukuda$^{(1)}$, 
Darryl Veitch$^{(3)}$,~\IEEEmembership{IEEE Fellow},\\
Kenjiro Cho$^{(4)}$,
Pierre Borgnat$^{(2)}$,~\IEEEmembership{IEEE Member},
Herwig Wendt$^{(5)}$,~\IEEEmembership{IEEE Member}\\
\makebox{} \\
$^{(1)}$  National Institute of Informatics/JFLI/Sokendai, Tokyo, Japan,  {{\tt kensuke@nii.ac.jp}},\\  
$^{(2)}$ Univ Lyon, Ens de Lyon, Univ Claude Bernard, CNRS,  Laboratoire de Physique, F-69342 Lyon, France {\tt patrice.abry@ens-lyon.fr, pierre.borgnat@ens-lyon.fr}, \\
$^{(3)}$ University of Technology Sydney, Australia, {\tt darryl.veitch@uts.edu.au}, \\
$^{(4)}$ IIJ Research Lab, Tokyo, Japan,  {\tt kjc@iijlab.net, romain@iij.ad.jp}, \\
$^{(5)}$ CNRS, IRIT, University of Toulouse, Toulouse, France,   {\tt herwig.wendt@irit.fr}
}

\maketitle


\begin{abstract}
    \let\thefootnote\relax\footnotetext{Work supported by French ANR grant MultiFracs ANR-16-CE33-0020. P.A. gratefully acknowledges the National Institute of Informatics for reccurrent \emph{Visiting Professor} funding.}
In the mid-90's, it was shown that the statistics of aggregated time series from Internet traffic departed from those of traditional short range dependent models, and were instead characterized by asymptotic self-similarity. 
Following this seminal contribution, over the years, many studies have investigated the existence and form of scaling in Internet traffic. 
This contribution aims first at presenting a methodology, combining multiscale analysis (wavelet and wavelet leaders) and random projections (or sketches), permitting a precise, efficient and robust characterization of scaling which is capable of seeing through non-stationary anomalies. 
Second, we apply the methodology to a data set spanning an unusually long period:  14 years, 
from the MAWI traffic archive, 
thereby allowing an in-depth longitudinal analysis of the form, nature and evolutions of scaling in Internet traffic, as well as network mechanisms producing them. 
We also study a separate 3-day long trace to obtain complementary insight into intra-day behavior. 
We find that a  biscaling (two ranges of independent scaling phenomena) regime is systematically observed: long-range dependence over the large scales, and multifractal-like scaling over the fine scales.
We quantify the actual scaling ranges precisely, verify to high accuracy the expected relationship between the long range dependent parameter and the heavy tail parameter of the flow size distribution, and relate fine scale multifractal scaling to typical IP packet inter-arrival and to round-trip time distributions. 
\end{abstract}

\section{Introduction}
\label{sec.motiv}

Statistical analysis and modelling of data traffic lies at the heart of traffic engineering activities for data networks including 
network design, management, control, security, and pricing.  
Surprisingly then, empirical measurements of computer network traffic did not appear until the early 1990's, making Internet modelling
in particular a somewhat young discipline.    In this contribution, we take advantage of an exceptional dataset which spans a good percentage of
this lifetime, to reexamine in depth one of the central features of Internet traffic -- scale invariance. 

\noindent {\bf Scale invariance in Internet traffic. } \quad
From the beginning, traffic processes, instead of being well described by models such as the Poisson process with its
independent inter-arrival times (IAT), or ARMA timeseries and other Markov processes with their richer but still short range
 (exponentially decaying) auto-correlation structures, were found to show significant \emph{burstiness} 
(strong irregularity along time) as well as slow, power-law decay of correlation \cite{Leland1994,Paxson1995,Erramilli1996,Willinger1997,AbryVeitch98}.
The latter phenomenon, referred to as \emph{asymptotic self-similarity} or as \emph{long range dependence} (LRD) \cite{Beran1994}, 
implies that no specific time scale or frequency plays a central role in the temporal dynamics of the data, 
a property also generically referred to as \emph{scale invariance}, \emph{scaling} or \emph{fractal} 
(for example see \cite{Park2000,ERVW2002}).
It was soon recognized that scaling had strong implications for networks due to its dramatic impact on queuing performance.
Indeed the discovery of `fractal traffic' stimulated much research in the queueing theory community 
(see \cite{fbmstorage,boxmasurvey,boxmacohensurvey}) which detailed the potentially severe performance penalties in 
terms of loss and delay of scaling arrival processes.

A natural question was that of the origin of scaling in traffic.
In the late 90's, a mathematical link was made relating a characteristic of underlying data objects to be served over the Internet, namely the heavy tail of their size distribution, to the LRD of aggregate traffic \cite{Willinger1997,onoffproof}.
To this day, this link remains the main framework used to explain the origin and nature of scaling in Internet traffic. 
Early empirical measurements of the tail index of file sizes qualitatively supported the finding 
as a realistic mechanism producing LRD in Internet traffic processes \cite{Leland1994,Crovella1997}.

The seminal observations 
described above drove a substantial research effort in the field over
the subsequent 20 years.
Evidence of scale invariance was reported continuously over this period for numerous different types of traffics and networks, 
e.g., \cite{Feldmann1998,AbryVeitch98,Feldmann1998b,Park2000,Veres2000,ERVW2002,abfrv02,hohn03,Karagiannis2004,pseudoMF,GLMT2005,CHMS2005,TerdikGyires} 
and works continue to appear \cite{roughanveitch07,MasugiTakuma2007,TerdikGyires2009,borgnat:infocom2009,lbfe2009,Park2010}. 
See also \cite{big_bib,willinger2002scaling} for early surveys. 
Despite being widely investigated however, there are a number of important challenges regarding Internet scaling which remain unresolved, even controversial.

\noindent {\bf Challenge 1:  Where is the scaling?} \quad
Although the existence of scaling phenomena in traffic, in particular LRD, is now essentially universally acknowledged, 
at a more detailed level important questions are routinely outstanding.
Scaling parameters, such as $H$ values and scaling ranges, measured on one network or one type of traffic 
often differ from those observed on others.
Even when measured on the same link over different days, or at different times within the day, 
scaling may be found to differ significantly.   
Differences are also found between links at the network edge compared to those in the core
and in large backbone networks (Tier-1 ISPs), where traffic volumes, multiplexing levels, and bandwidths, are all higher.
Finally, measurements typically consist of a mixture of \emph{normal} background traffic, corresponding to a base load of legitimate traffic, with sporadic \emph{anomalous} traffics, be they legitimate such as flashcrowds, or malicious such 
as aggressive Denial-of-Service Attacks \cite{CHMS2005,borgnat:infocom2009}.
This results in a paradoxical situation where, despite having far more data available than in most other applications, pernicious non-stationarities, which cannot be eliminated by simple time averages, induce a lack of statistical robustness and reproducibility of conclusions. 
These considerations can even lead to doubts as to the very existence of scale invariance, seen instead
 as a spurious empirical observation produced by non stationarities.

\noindent {\bf Challenge 2: Is there scaling beyond LRD?}  \ 
LRD describes scaling in the auto-correlation of the data and thus only concerns 2nd order statistics.
It therefore neglects the impact of departures from Gaussianity, a much debated issue \cite{Feldmann1998b, Kilpi2002,Zhang2003}.
To model potentially richer scaling involving higher order statistics, and the full dependence structure including departure from Gaussianity, the multifractal paradigm was put forward \cite{TTW97,Feldmann1998b,Walter_MF_Allerton,Feldmann1998,Soucy99,riedi,sarvotham2001connection,abfrv02}. 
Multifractal models explicitly designed for Internet traffic were proposed in \cite{riedi,Kent99,Soucy99}, while the impact of multifractality on performance was investigated in \cite{cascadeperf}, thus showing its practical importance.
Deciding whether Internet traffic could be multifractal or \emph{simply} self-similar became important as the former implies significant departures from Gaussianity as well as the presence of underlying cascade-like multiplicative mechanisms \cite{Jaffard2015}.
Together with discussions of its possible origins, the existence of multifractality in traffic has been the subject of numerous investigations, with sometimes contradictory conclusions \cite{TTW97,Feldmann1998b,Walter_MF_Allerton,Feldmann1998,pseudoMF,MasugiTakuma2007,LOISEAU:2010:A,Jaffard2015}.
For example \cite{pseudoMF} points out that among the initial papers examining the issue of multifractal scaling in 
traffic, conclusions which appeared at times at odds were in fact not, as they were made in relation to different scale ranges.
More generally, assessing both the existence of scaling and its nature brings into focus the importance of the selection of the range of time scales where scaling properties are observed and analysed, an issue whose importance is often overlooked and/or underestimated (see a contrario \cite{hohn03,Jaffard2015,LOISEAU:2010:A}).

\noindent {\bf Challenge 3: Is scaling here to stay?}  \quad
The Internet has evolved rapidly since its creation, and it is commonly accepted that this will continue as new services and applications, business models and regulation regimes, protocols and control plane paradigms, 
as well as hardware and software, evolve. 
For example clearly the Internet today conveys much larger volumes of traffic, 
at far higher bandwidth, than in the year 2000.
More recent examples include the rise of traffic from social media such as \emph{twitter}, and the 
democratization of protocols and the redesign of routing enabled by Software Defined Networking (SDN).
This has lead some to argue that the statistical properties of Internet traffic in the modern era should be very different 
from that of the early days (barely 20 years ago).
Notably, essentially relying on a \emph{Central Limit Theorem} argument, these analyses suggest that traffic 
statistics will return to being Gaussian and Poisson-like, implying the irrelevance or disappearance of scale invariance
(see interesting discussions and analyses in \cite{Cao2001,Cao2001b,Karagiannis2004,TerdikGyires,TerdikGyires2009}).

%

\noindent {\bf Challenge 4: Is scaling an Internet invariant?}  \quad
Studies of Internet traffic scaling reported in the literature typically concentrate on one or a small number of traces, 
collected at specific times, often with a focus on the latest killer application or a fascination for previously unseen phenomena.
However, statistical analyses solid enough to address the challenges outlined above can only be achieved through 
longitudinal studies, making use of a large data corpus, collected along several years, as is the case for example in \cite{Claffy1994,TerdikGyires, Fomenkov2004,Karagiannis2004,JD2005,pseudoMF, CHMS2005, Allman2007,borgnat:infocom2009,grm,Park2010}. 
There exist only a few trace repositories where such a large corpus of data are available: 
\href{http://ita. ee. lbl. gov}{\rm Bell labs}, 
\href{http://wand.cs.waikato.ac.nz/wand/wits/}{\rm WAND}, 
\href{http://www.caida.org/analysis/workload/oc48/}{\rm CAIDA MFN Network}, 
\href{http://mawi.wide.ad.jp/mawi/}{\rm MAWI}.



\begin{figure*}
\includegraphics[width=\textwidth]{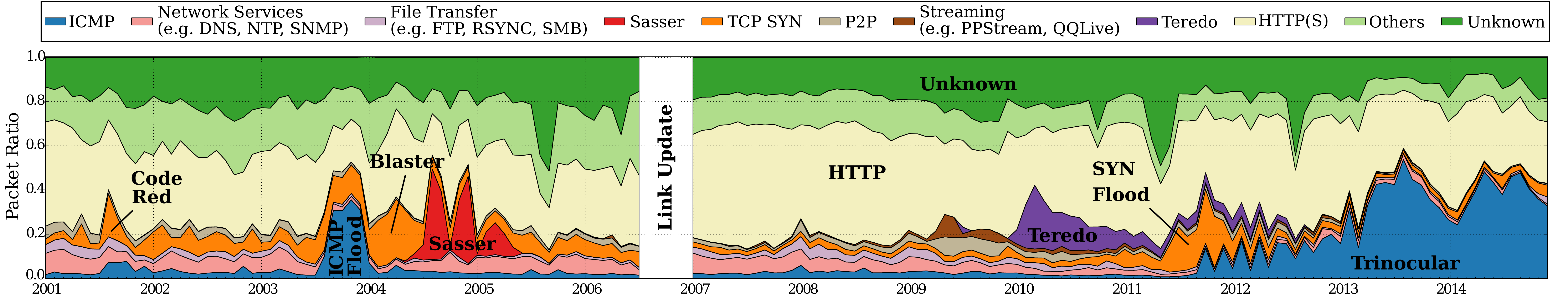}
\vskip-2mm 
 \caption{{\bf MAWI Traffic.} Application breakdown, from Jan. 2001 to Dec. 2014 (monthly level aggregation).}
 \label{fig:mawiBreakdown}
\end{figure*}

\noindent {\bf Goals, contributions and outline.} \quad 
Today, 20 years after the original reports of long memory in data traffic \cite{Leland1994},  Internet scaling behavior
is no longer a hot topic.  Nonetheless, the challenges described above remain, making the development of a definitive
understanding of traffic scaling, and the goal of definitive and widely accepted traffic models, no closer to fruition now than 10 years ago.  

We contend that the time is right to revisit this topic, because of the conjunction of two opportunities. 
First, the MAWI 
repository, which has been collecting traces daily since 2001, constitutes an exceptionally rich dataset, encompassing a diversity of applications and network conditions, including the presence of major known anomalies with global impact or local ones, congestion periods, link upgrades and network reconfigurations.
This provides a unique opportunity to perform a longitudinal study of the statistical scaling properties of Internet traffic,
 over 14 years, an exceptionally long period of time in relation to the lifetime of the field itself.
Second, there is an opportunity thanks to the greater maturity of statistical methods, compared to those used before. 
We combine the use of \emph{random projections} or \emph{sketches} to provide robustness against the debilitating issue of anomalies which fatally distort statistical analysis, and use \emph{wavelet leaders} for the precise assessment of scaling properties, in particular to handle the difficult issue of the empirical measurement of multifractality.  
These are now understood to be far superior to other approaches, including normal wavelet analysis, for this purpose~\cite{WENDT:2007:E,Jaffard2015}. 

Our goals are also twofold, based on combining the above two opportunities.
First, to apply the new tools to the unique dataset, in order to obtain reliable longitudinal results, and therefore to
meaningfully contribute to the resolution of the challenges described above.  
As part of this, we provide some of the most robust evidence ever presented for the presence of various kinds
of scaling, and in particular, a high quality validation of the link described above between heavy tailed sources and LRD.
Second, we mine the MAWI repository to provide elements toward responses to the challenges raised above and particularly toward a greater understanding of mechanisms underlying
scaling at `small' scales. 

Section \ref{sec:data} describes the MAWI archive and the datasets we use. 
The theoretical and practical methodology to study scaling in Internet traffic (sketches and wavelet leaders) is detailed in Section~\ref{sec:scaling}.
Applying these methodological tools to data from the MAWI repository then allows: 
i) to robustly assess the existence of different scaling properties in traffic, with discussions of the different ranges of time scales involved (Section~\ref{sec:RB});
ii) to quantify short term intra-day variations of scaling properties as well as long term evolution over 14 years 
(Section~\ref{sec:evol}); and
iii) to characterize the nature of scaling (LRD versus multifractality) both in the coarse and fine scale ranges, and to investigate quantitatively and qualitatively the mechanisms potentially producing such scaling (Section~\ref{sec:nature}). 
\vspace{-2mm}

\section{Data}  
\label{sec:data}

\noindent \textbf{The MAWI repository.}
The MAWI archive \cite{Cho2000a} is an on-going collection of Internet traffic traces, captured within the \textit{WIDE} backbone network (AS2500) that connects Japanese universities and research institutes to the Internet. 
Each trace consists of IP level traffic observed daily from 14:00 to 14:15 (Japanese Standard Time) at a vantage point within WIDE, and includes each IP packet, its MAC header, and an \emph{ntpd} timestamp.
Anonymized versions of the traces (with garbled IP addresses and with transport layer payload removed), are 
made publicly available at \href{http://mawi.wide.ad.jp/mawi/}{\rm http://mawi.wide.ad.jp/}.

As WIDE peers with all major domestic ASes, it used to mainly carry trans-Pacific traffic. 
However, as the global network topologies become less US-centric and content providers start operating their own networks and become less dependent on major ISPs, it now carries a rich traffic mix including academic and commercial traffic.
Consequently MAWI traces, which typically contain several 100k IP addresses, capture diverse behavior, as summarized by the breakdown of traffic types over 14 years shown in Fig.~\ref{fig:mawiBreakdown}, obtained with the traffic classifier \emph{libprotoident}.
Although largely dominated by HTTP, the traffic composition is markedly influenced by unusual events.
Some of these are global, for example, \emph{Code Red}, \emph{Blaster} and \emph{Sasser} are worms 
that infamously disrupted Internet traffic worldwide \cite{Allman2007}.  
Of these, Sasser (2005) impacted MAWI traffic the most, accounting for 68\% of packets at its peak.
Conversely, the ICMP traffic surge in 2003, and the SYN Flood in 2012, are more local in nature,
each revealing attacks on targets within WIDE that lasted several months.
The period covered by Fig.~\ref{fig:mawiBreakdown} also includes congestion periods (from 2003 to 2006) \cite{borgnat:infocom2009}, and changes in routing policy (2004).
MAWI traffic has also been significantly altered by temporary deployments or research experiments.
The surge of Teredo traffic in 2010 is due to the IPv6 traffic temporarily tunneled by the \emph{Tokyo6to4} project, and the increase of ICMP traffic from September 2011 is caused by \emph{Trinocular}, 
an experimental outage-detection system that actively probes Internet hosts \cite{trinocular,fontugne2015icassp}.

\noindent \textbf{Datasets.} 
The traces used here were taken from those collected daily from \textit{samplepoint-B} within WIDE until its 
decommissioning in July 2006, and from \textit{samplepoint-F} from Oct. 2006 onwards.  
\textit{Samplepoint-F} is on the same MAWI router as \textit{samplepoint-B}, but connected to a new link following a network upgrade and reconfiguration.
These links have capacity, respectively, of 100Mpbs with 18Mbps Committed Access Rate (CAR, an average bandwidth limit), and 1Gbps  with a CAR of 150Mbps.
We extract from each packet record the packet size, timestamp, and when needed, 
a standard header 5-tuple
(IP address and port number for source and destination, and IP protocol carried (TCP, UDP or ICMP))
used to construct flow-ids.
We examine two data sets.


\noindent {\bf DataSet I (Longitudinal): 15-min traces 2001-2014. } \\
A total of 1176 standard traces taken from the first 7 days of each month from Jan. 2001 to Dec. 2014.

\noindent {\bf DataSet II (Intra-day):  3-day trace 2013.} \\
To allow a study of intra-day variations, a special 3-Day long trace was measured at \textit{Samplepoint-F}  over June 25-27, 2013.
This trace, containing  a large variety of applications, features a strong diurnal cycle where the packet rate during working hours is about twice that at night, yielding a change in the typical packet inter-arrival time from 0.01ms to 0.023ms.
\vspace{-2mm}


\section{Scaling analysis - Theory \& Methodology}  
\label{sec:scaling}

\subsection{Scaling and multifractal analysis} 
\label{sec-si}

The goal of this section is to briefly introduce both the analysis tools, (wavelet coefficients \cite{AbryVeitch98,va99,va01,abfrv02,Karagiannis2004}, wavelet leaders \cite{WENDT:2007:E,Jaffard2015}) and stochastic models (LRD \cite{Beran1994,Leland1994} and multifractality \cite{WENDT:2007:E,riedi,Feldmann1998b,Jaffard2015}), that are essential to a discussion of scale invariance in Internet traffic.

\subsubsection{LRD and wavelet coefficients} 

Let $\psi$ denote a \emph{mother wavelet}, characterized by an integer $N_\psi>0$, defined as 
$\int_\real t^k \psi(t) dt \equiv 0$ $\forall n = 0, \ldots, N_\psi\!-\!1$,  and $ \int_\real t^{N_\psi} \psi(t) dt \neq 0$, 
known as the number of vanishing moments.
{The ($L^1$-normalized) discrete wavelet transform coefficients of a process $X$ are defined as
$d_X(j,k) =  \langle \psi_{j,k}|X\rangle$, with  $\{ \psi_{j,k}(t) = 2^{-j} \psi(2^{-j}t-k) \}_{(j,k) \in \N^2}$. 
For a detailed introduction to wavelet transforms see \cite{Mallat1998}.


When $X$ is a process which exhibits second order scaling 
the time average of the squared wavelet coefficients behave as a power law with respect to the analysis time scale $a=2^j$,  
\begin{equation}
\label{equ-scaling}
	S_d(j)  \equiv {1 \over n_j} \sum_{k=1}^{n_j} d^2_X(j,k) \simeq C 2^{j(2H-2)}, 
\end{equation}
over a range of scales,  $2^{j_1} \leq 2^j \leq 2^{j_2}  $, with $\frac{2^{j_2} }{2^{j_1} } \gg 1$,
where $n_j$ denotes the number of $d_X(j,k)$ actually available at scale $2^j$) \cite{AbryVeitch98}.
The scaling exponent $2H-2$ is driven by the \emph{Hurst parameter} $H$, that usually takes values in $H\in(0,1)$. 
For example in the case of a stationary $X$ with LRD, $H\in(0.5,1)$ and $2^{j_{\max}}=\infty$, 
and the resulting slow decay of correlations over all time poses significant statistical challenges, 
which motivates the recourse to wavelet analysis \cite{abry98b,AbryVeitch98,va99,va01}.

\miniskip
\subsubsection{Multifractality and wavelet leaders}  
 
Eq.~(\ref{equ-scaling}) is related to the (algebraic or power-law decay of the) correlation function of $X$ only.
Multifractal analysis describes the statistical properties of data not just at second order but at arbitrary $q$th order, for processes where Eq.~(\ref{equ-scaling}) can be nominally generalized to  
$ {1 \over n_j} \sum_{k=1}^{n_j} |d_X(j,k)|^q\simeq C 2^{j\zeta(q)}$. 
Multifractal analysis assesses whether 
{$\zeta(q) \equiv q(H-1)$,
where $H$ alone controls scaling at all orders, 
or departs from this linearity in $q$, revealing richer temporal dependencies, referred to as \emph{multifractality}.

It is now theoretically well-grounded and practically well-documented 
\cite{WENDT:2007:E,Jaffard2015} 
that correctly assessing the linearity of $\zeta(q)$ requires wavelet coefficients to be replaced with \textit{wavelet leaders}.
Let $\lambda_{j,k}=[k2^j,(k+1)2^j)$ denote the dyadic interval of size $2^j$ centered at $k 2^j$, 
and $\threed$ the union of $\lambda_{j,k}$ with its neighbors:  
$\threed=\lambda_{j,k-1}\cup \lambda_{j,k}\cup\lambda_{j,k+1}$.
The \textit{wavelet leader} $L_X(j,\kb)$ is the largest wavelet coefficient over all finer scales $j'<j$
within $\threed$: $L_X(j,\kb):=  \sup_{\lambda'\subset\threed} 2^{j'}\vert d_X(\lambda')\vert$, with factors $2^{j'}$ and $2^{-j}$ used here to compensate for insufficient regularity of $X$, cf. \cite{WENDT:2007:E,Jaffard2015}. 

The exponent $\zeta(q)$ can now be measured via
\vskip-2mm
\begin{equation}
\label{eq:se}
	S_L(j,q) \equiv \frac{1}{n_j} \sum_{k} L_X(j,k)^q \simeq S_{0}(q){2}^{j\zeta(q) }. 
\end{equation}
\vskip-2mm
The Legendre transform ${\cal L} (h)$ of $\zeta(q)$ provides an estimation of the 
\emph{multifractal spectrum} which defines the multifractal properties of $X$
via the fluctuations of its pointwise H\"older exponents $h$, cf. \cite{WENDT:2007:E,Jaffard2015}.
For further theoretical details on multifractal analysis and wavelet {\r l}eader formalism, the reader is referred to, e.g., \cite{WENDT:2007:E}. 
For our purposes here however, we focus on characterising $\zeta(q)$, specifically its linearity. 
It is advantageous to do so indirectly, via the cumulants  $C_p(j)$ of order $p$ of $\ln  L_X(j,\kb)$.
It can be shown (\cite{WENDT:2007:E}), that when $X$ has multifractal properties, the $C_p(j)$ take the explicit form
\begin{equation}
\label{equ:cum}
	C_p(j) = c^0_p + c_p \ln 2^j
\end{equation}
where the $c_p $ can be directly related to the scaling exponents $\zeta(q) \simeq c_1 q + c_2 q ^2/2 + \ldots$, and hence to the multifractal spectrum.
The first two cumulants are sufficient for our purposes as a measurement of 
$c_2 < 0$ implies nonlinearity in $\zeta(q)$ and hence multifractality.
 It amounts to assuming a parabolic multifractal spectrum 
${\cal L} (h)  \simeq 1 + (h-c_1)^2/(2c_2)$, 
where $c_1$ controls the position of the maximum of  ${\cal L} (h)$ while $c_2$ quantifies its width (see \cite{WENDT:2007:E,Jaffard2015} for details). 
When $c_2 \equiv 0$, one has $H \equiv c_1$, otherwise, one approximately obtains $H \simeq c_1 + c_2 $ for times series with true scaling properties.

%
%
\subsubsection{Scaling range and estimation of scaling exponents}
\label{sec:estim}

The estimation of $H$ is wavelet based, performed by linear regression of $\log_2 S_d(j)$ against $j=\log_2 2^j$.
The estimation of $c_p$ is wavelet leader based, performed by regressing $ C_p(j)$ against $j$.
A plot of $\log_2 S_d(j)$, $C_1(j)$ or $C_2(j)$ as a function of log-scale $j$ is referred to as a \emph{Logscale Diagram} (LD).
By  \emph{best fits}, we mean the estimation procedures extensively detailed in \cite{va01,WENDT:2007:E}.
Whereas $\log_2 S_d(j)$ and $C_1(j) $ (estimating respectively $\log_2 \E[D^2_X(j,k)]$ and $\E[\ln L_X(j,k)]$), and thus $H$ and $c_1$ are mainly associated to the 2nd-order statistics of $X$, $C_2(j)$ (estimating $\var[\ln L_X(j,k)]$) and hence $c_2$, conveys information beyond correlation.
The crucial prior step to estimation is to carefully examine LDs to determine the range of scales $2^{j_1} \leq a \leq 2^{j_2}$ over which the regression is performed, 
either by manual inspection or using goodness-of-fit tests.
Thus, LDs plots, estimation and tests are assessed by time-scale domain bootstrap based procedures (cf. \cite{va99,va01,WENDT:2007:E,Leonarduzzi2014}).
Practical scaling and multifractal analyses were conducted using a toolbox designed by ourselves and publicly \href{http://www.irit.fr/~Herwig.Wendt/software.html#wlbmf}{\rm available}.
\vspace{-6mm}

\subsection{Packets versus bytes and aggregation procedure}
\vskip-1mm
In nature, Internet traffic consists of a flow of IP Packets and could thus \emph{naturally} be modeled as a (marked) point process.
However, analyzing such point processes would require massive memory and computational capacities.
It is thus often preferred to analyse aggregated time series, consisting of the count of packets (or bytes) within bins of size $\Delta_0$, the choice of which being often considered arbitrary and sometimes controversial. 
However, a wavelet transform can be considered per se as an aggregation procedure thus making the actual choice of $\Delta_0 $ much less crucial, as it does not imply a narrow analysis at that 
scale, but rather an analysis over all scales $a\ge\Delta_0$, cf. \cite{abry98b}.
Wavelet analysis does however require an initialising projection into an 
\emph{approximation space} at the initial scale $\Delta_0$, which we approximate here (with negligible error)
by a simple packet count in each `bin' of width $\Delta_0$.

We study the packet arrival times, referred to as the packet arrival process $X_{\Delta_0}(t)$. 
We do not consider the IP byte arrival process, both for space reasons and because prior work
\cite{Dewaele2007,borgnat:infocom2009} suggests that the main features of the two are the same.

For point processes, scaling cannot exist at scales finer than the typical \emph{inter-arrival time} (IAT), 
 $\tau$.
To permit an analysis of the finest meaningful scales, it is thus natural to choose $ \Delta_0 \simeq \tau$.
 We select  $\Delta_0 \equiv 2^{-3} = 0.125$ms which is of the order of the median IAT for both 15-min and 3-day traces. 
Time scales are normalized with respect to $\Delta_0$, that is  $ \Delta_\tau = \Delta_0 2^{j_\tau}$. 
Hence scale $a= 1=2^0$ (octave $j=0$) refers to $\Delta_0$.

\subsection{Random projections (or sketch procedure)}  
\label{sec:Sketch}
\vskip-1mm

\subsubsection{Robustness from averages}


As discussed above, the variety in network topologies, volumes or the nature of the traffic itself, often leads to a failure of reproducibility in its statistical analysis. 
Combined with such diversity, the existence of non-stationary anomalous traffic superimposed onto normal traffic, 
whether malicious or not, essentially precludes the use of time averages to overcome this lack of statistical robustness.
However, compared to data from other application fields, Internet traffic has the particularity that, beyond the aggregated time series $X_{\Delta_0}(t)$ itself, assembled from IP Pkt timestamps, the extra 5-tuple available for each packet conveys valuable information that can be used to robustify statistical analysis. 
We follow the approach pioneered in \cite{Krishamurthy2003,Muthukrishnan2003}, and elaborate on the methodology developed in \cite{borgnat:infocom2009}, to use random projections to circumvent this core difficulty.

\subsubsection{Random projection}
\label{sec-RP}

A random projection (or sketch procedure) \cite{Muthukrishnan2003,Krishamurthy2003} relies on the use of a $k$-universal hash function $h$ \cite{thorup:siam04}, applied to an IP Pkt attribute $A$, chosen by practitioners, that defines a notion of \textit{flow}, and taking values in an alphabet of size $2^M$. 
A sketch procedure thus splits an original IP trace $X$ into $2^M$ sub-traces, 
$X^{(m)}_{\Delta_0}, m=1,..., 2^M$,  
each consisting of all packets with identical sketch output $ h(A)$,  thus preserving flow structure (packets belonging to a same flow are assigned to the same sub-trace).
The intuition here is that when there are no anomalies, random projections amount to creating surrogate traces which can be expected to be only weakly dependent if $M$ is large, and that are statistically equivalent to each other up to a multiplicative factor.
Conversely, when present,  anomalous flows are likely to be concentrated in a subset of the sketches.
Robust estimation then stems from using a median procedure 
across sketch outputs, thus providing a reference for normal traffic that shows little sensitivity to the anomalies. 


\subsubsection{Hash key}
\label{sec-sk}

Selecting the hash key for defining flows is important, as different choices will lead to different sub-traces. 
We used Source IP and Destination IP addresses as obvious choices, and found equivalent conclusions in terms of the statistical characterization of scale invariance in Internet traffic, even though certain types of anomalies are missed using either hash key. 
Yet, the ultimate goal is not anomaly detection, but rather robustness of the median statistics. 
Because both keys lead to equivalent statistical description, results reported below were obtained using Source IP address as the hash key.

\begin{figure*}
\centerline{
   \includegraphics[width=.33\linewidth]{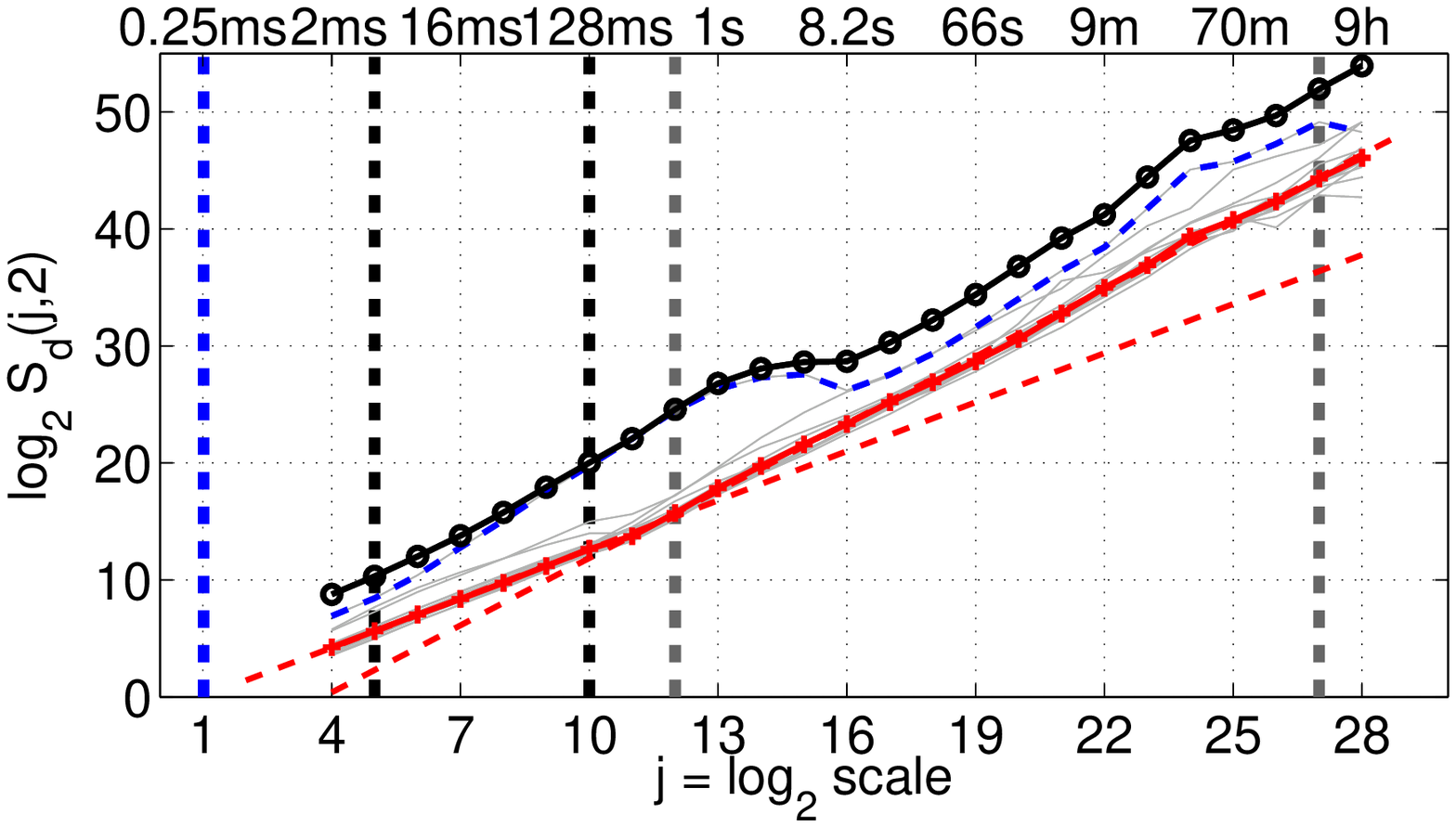}
  \includegraphics[width=.33\linewidth]{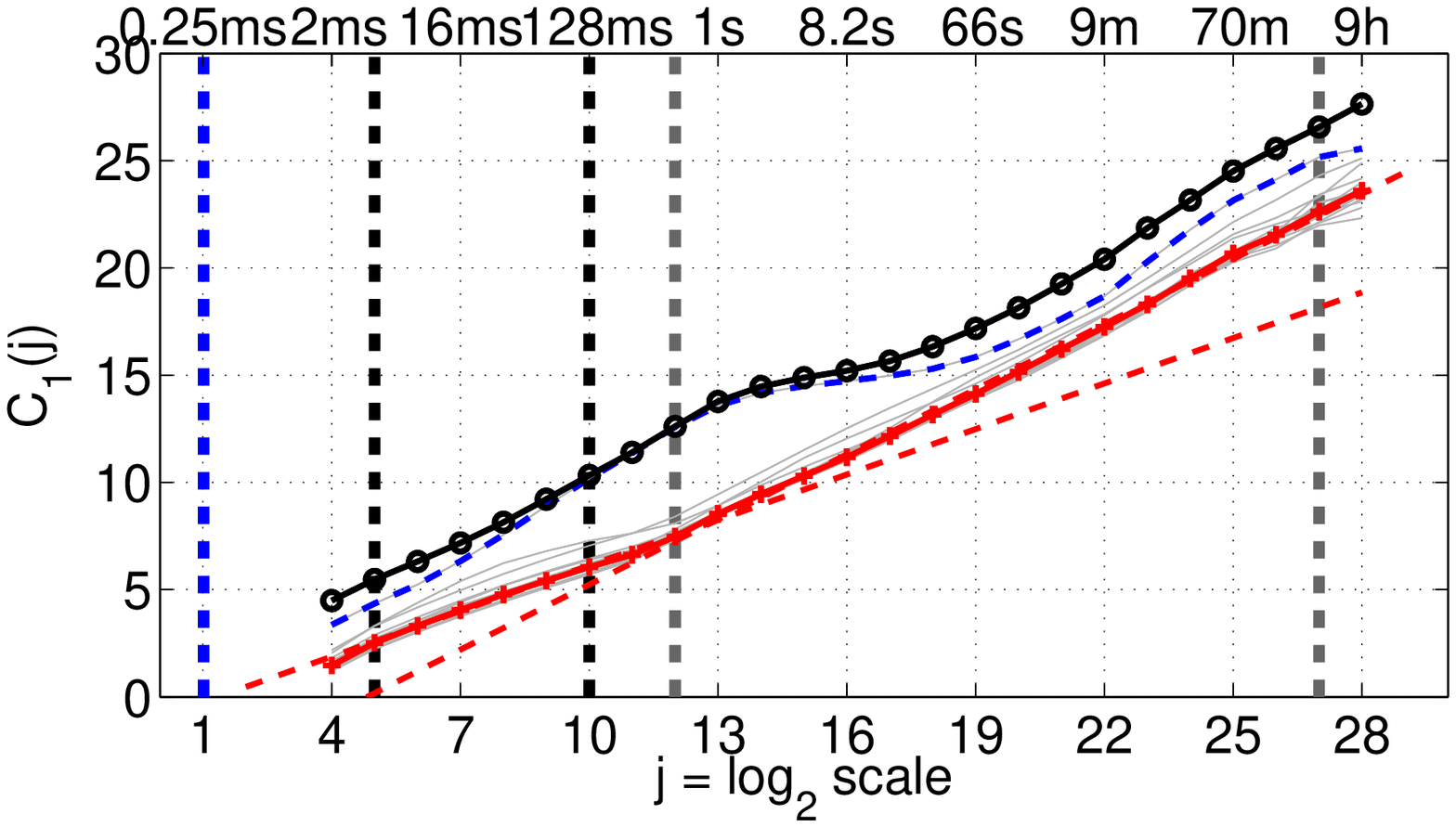}
  \includegraphics[width=.33\linewidth]{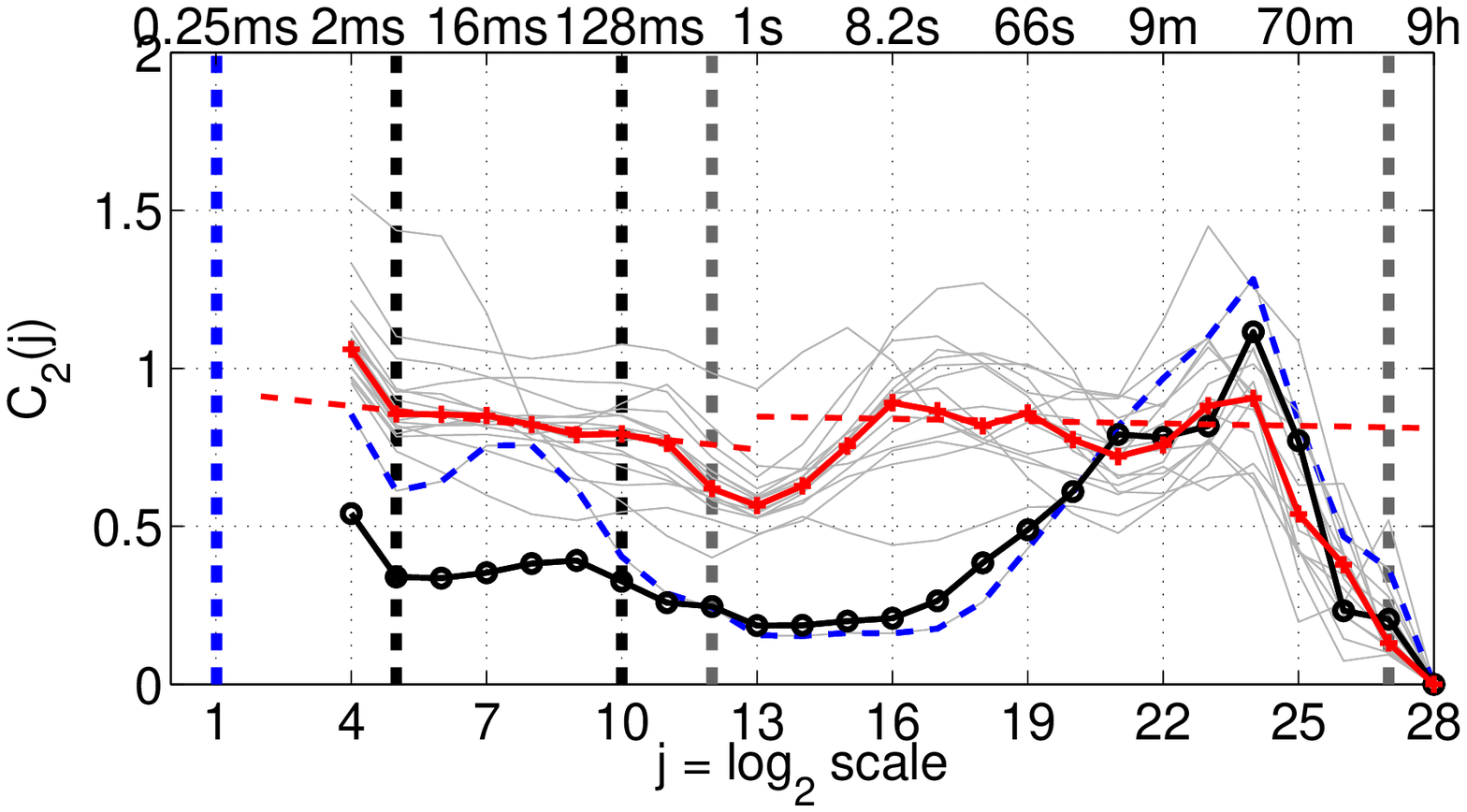}
  }
  \centerline{
 \includegraphics[width=0.33\linewidth]{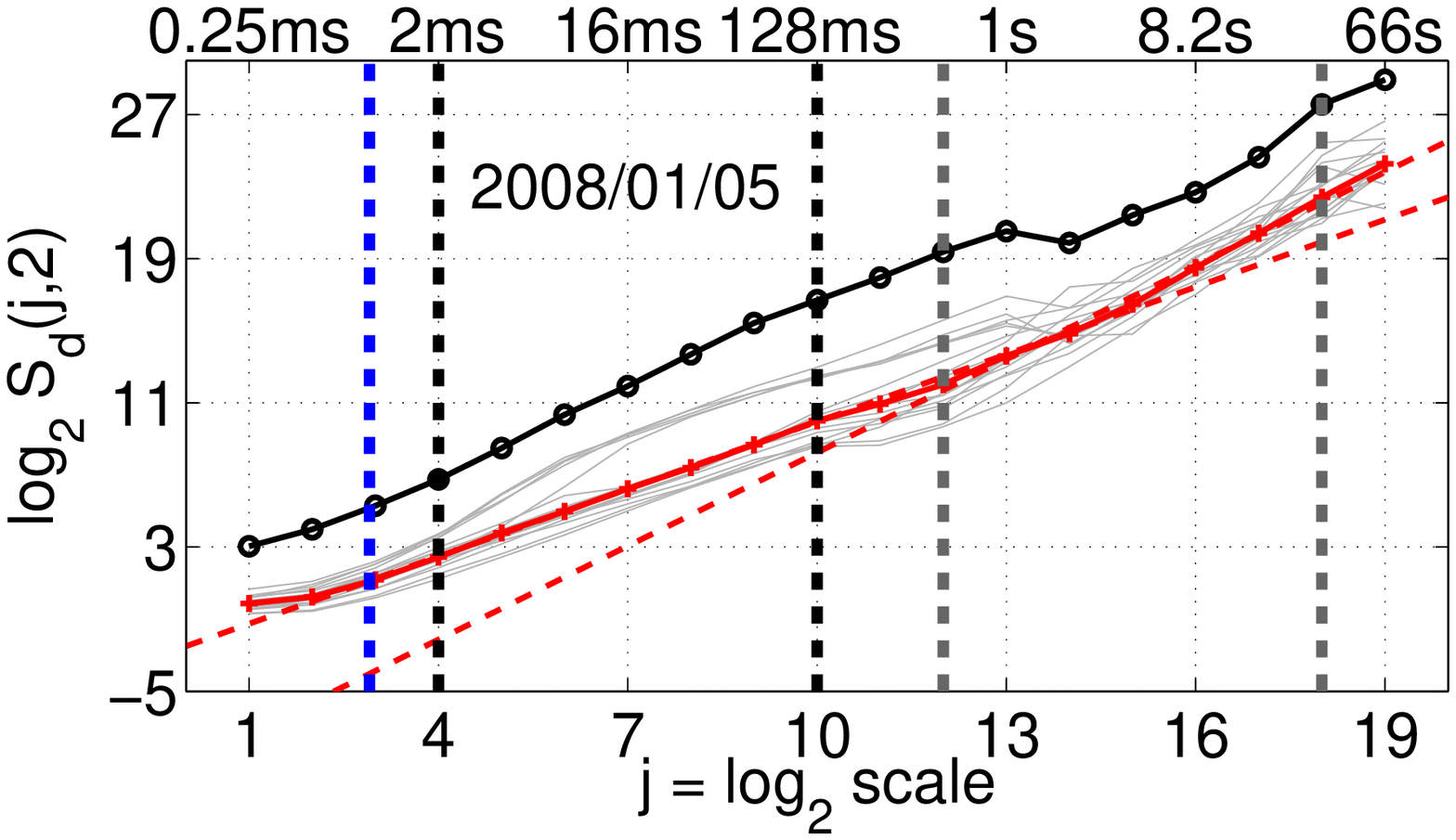}
 \includegraphics[width=0.33\linewidth]{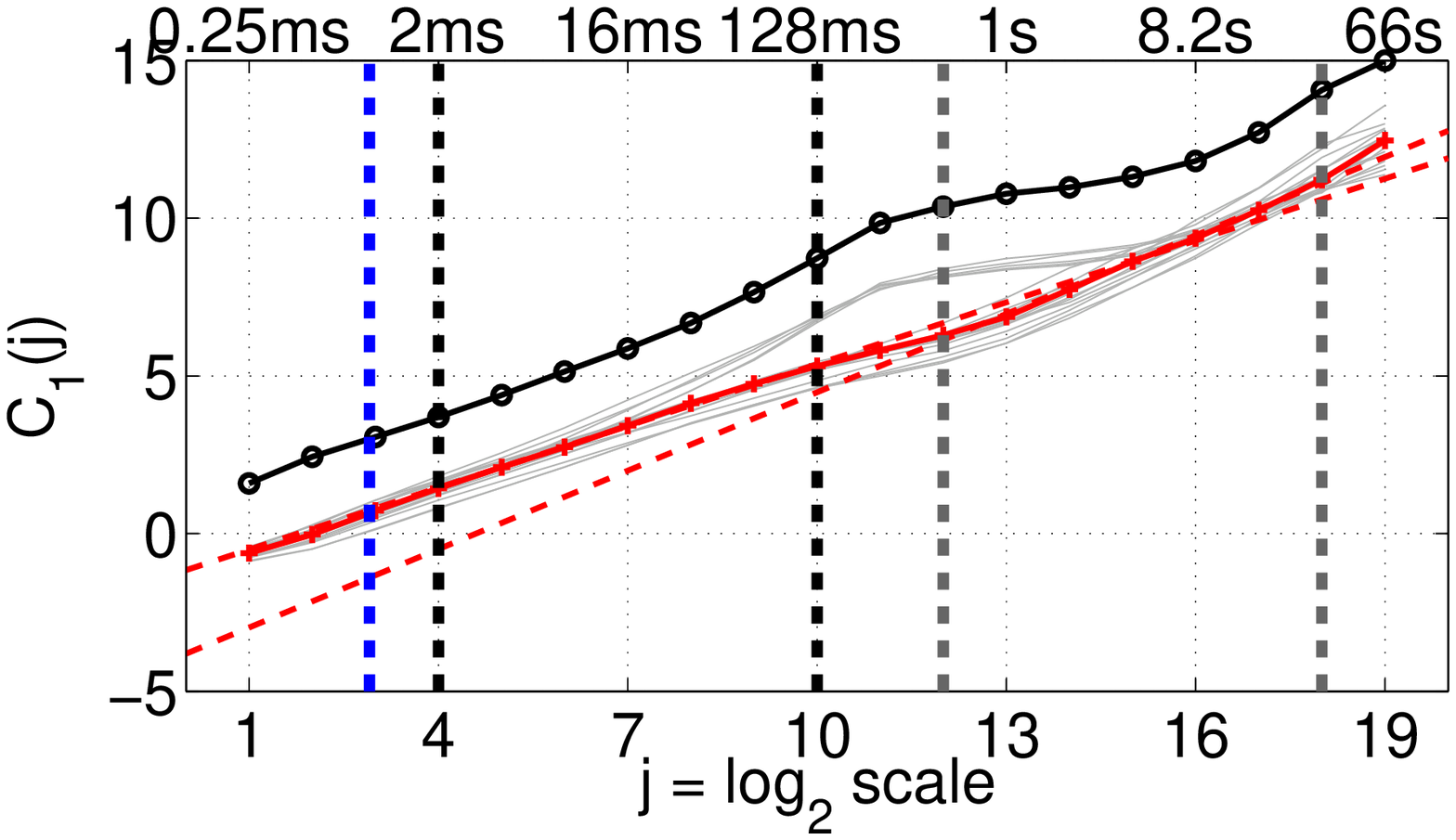}
 \includegraphics[width=0.33\linewidth]{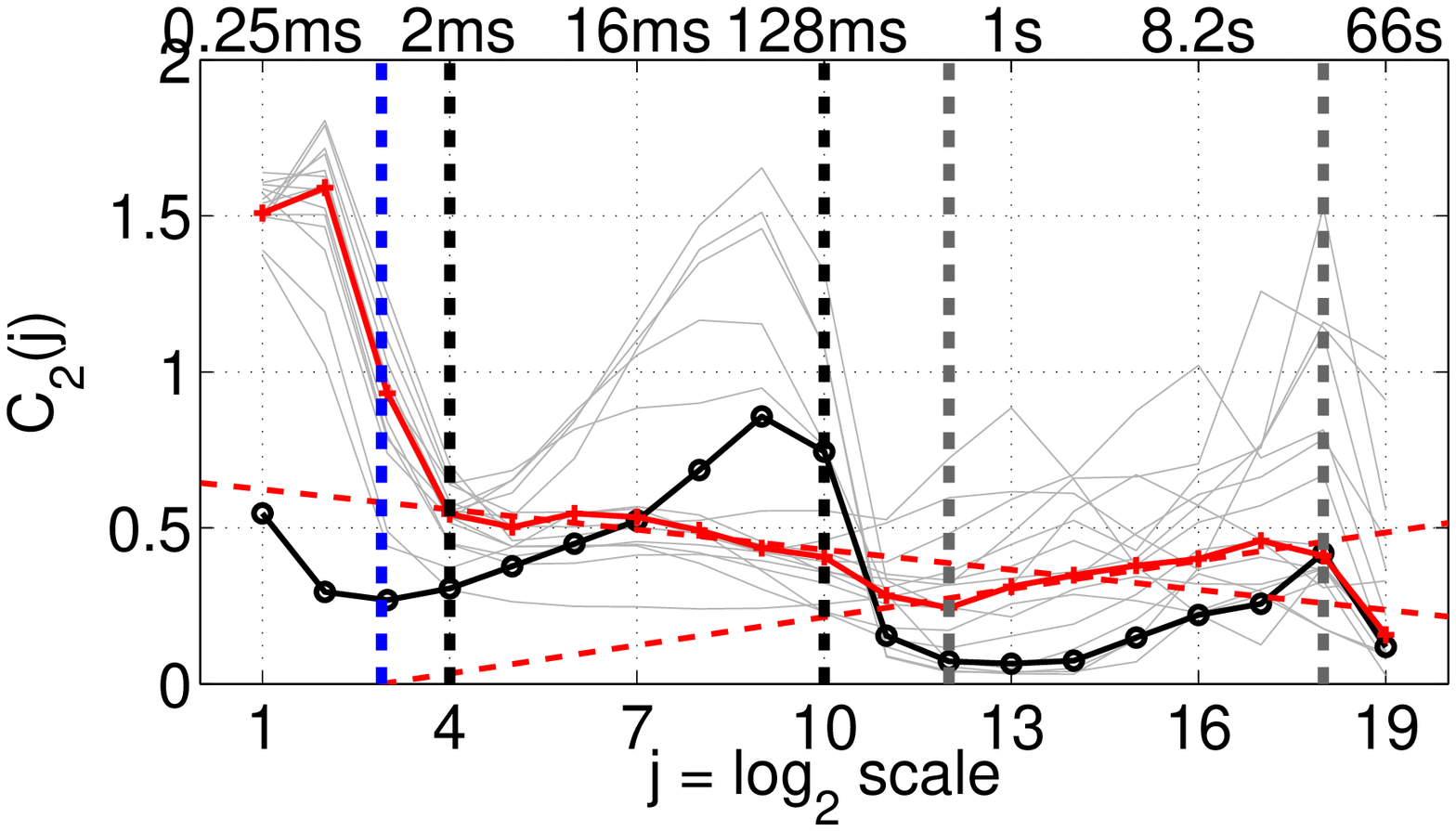}
}
  \vskip-2mm
\caption{\label{fig:LD3Da}{\bf LDs.}  
From left to right, LDs for the statistics $\log_2 S_d(j)$, $C_1(j)$ and $C_2(j)$.
Each plot shows Global-LDs (black), Sketch-LDs (light grey), Median-LD (red). 
Global-LDs are dominated by the Sketch-LD concentrating anomalous traffic (dashed blue). 
The dashed vertical lines mark respectively the typical IAT time scale $ j^M_\tau $ (blue), and the FS (black) and CS (dark grey) scaling ranges $(j_1,j_2)$. 
{\it Best fits} of Median-LDs are shown (dashed-red lines) over both FS and CS.
Top: for the 3-day trace. Bottom for one arbitrarily chosen 15-min trace.}
\end{figure*}

\section{Robust biscaling in MAWI traffic} 
\label{sec:RB}

\subsection{Robustness via random projections}
\label{sec:R}


The scale invariance analysis described in Section~\ref{sec-si} produces systematically 3 LD plots, $\log_2 S_d(j)$, $C_1(j)$ and $C_2(j)$.
When applied to the full trace, they are referred to as \textit{Global-LD}s, and when applied to a sub-trace, as \textit{Sketch-LD}s.
Over a set of M Sketch-LDs from a given trace, \textit{Median-LD} is obtained as a pointwise median of $M$ values for each $j$.
Fig.~\ref{fig:LD3Da} compares, for the 3-day trace and for one 15-min trace, separately for $\log_2 S_d(j)$, $C_1(j)$ and $C_2(j)$, the Global-LDs,  the individual Sketch-LDs, and the Median-LD.
Equivalent plots for each 15-min trace are available online\footnote{http://romain.iijlab.net/internetScaling/results.html}.

\subsubsection{Robustness}
\label{sec:RP}

{Examining all LD plots leads to conclude that:}
i) The shapes of Global-LDs are clearly driven by a particular sketch output, whose shape differs from that of the majority of the Sketch-LDs~;
ii) As expected, the shapes of the majority of Sketch-LDs are close to that of the Median-LD~;
iii) Global-LDs do not show clearly identifiable scaling ranges where alignment is seen, corresponding to power-law behavior (scaling), whereas the Median-LDs do~;
iv) Global-LDs vary markedly over days, 
whereas Median-LDs remain fairly comparable~; 
v) Global-LDs show intraday diurnal cycles, 
whereas Median-LDs remain fairly constant.

Making use of the sketch-based detection procedure, proposed in \cite{Dewaele2007}, to identify the IP addresses involved in traffic anomalies, enabled us to determine that the sketch output driving the global-LDs in Fig.~\ref{fig:LD3Da} 
is dominated by a high volume component of ICMP traffic involving a single source IP address. 
It in fact corresponds to probing data from \emph{Trinocular} (cf. Section~\ref{sec:data}).
In a similar way, for the 15min-traces, it was verified that the Sketch-LDs that essentially drive the global-LDs can be quasi-systematically associated to anomalous traffic. 
It is important to note that there is almost not a single day free of significant anomalies 
(see \cite{Dewaele2007,borgnat:infocom2009}). 

\subsubsection{The number of sketches}
\label{sec:SO}

The choice of the number $2^M$ of sketch outputs results from a trade-off: 
robustness against anomalous behaviors increases with higher $M$ where it is more likely that anomalies will be isolated in a small number of outlier subtraces, while the remainder of sketch outputs can be regarded as surrogate traces acting as a large number of 
copies of traffic with equivalent statistical properties.
Increasing $M$ however also results in subtraces whose statistics may diverge from those of the background traffic as a whole.  
In particular, it results in an increase in the IAT  of each subtrace: 
$\Delta^M_\tau = 2^M \Delta_\tau = \Delta_0 2^{j^M_\tau}$, with $j^M_\tau = j_\tau+M$, 
and so impairs the statistical analysis of background traffic at the finest scales. 
Given that the analysis of scaling requires access to the widest possible range of scales, 
we select  $M = 4$ to keep $j^M_\tau $ reasonably low. 

\subsubsection{Partial conclusion 1}


Global-LDs are strongly impacted by anomalous traffics, and are likely to change often.
They can therefore not be analyzed reliably without an ability to filter out anomalies, which the random projection and median-sketch procedure provides.  
Median-LDs characterize the statistics and scaling properties of the traffic with many anomalies rejected, and constitute a de facto legitimate background traffic.
\vspace{-3mm}

\subsection{Biscaling} 

\subsubsection{Two scaling ranges} 
\vspace{-1mm}

The analysis of the Median-LDs 
leads to a significant, and robust observation: 
The statistical signature of background traffic does not consist of a single scaling range across all scales but rather of two separate scaling ranges.
This is consistently observed across the 
all 15-min traces but for exceptions, occurring in less than 1\% of cases, because of data quality issues)
as well as for the 3-day trace.
This clearly implies that the packet arrival process is not describable by a single scaling 
mechanism acting over all available scales, 
ranging from milliseconds to hours, but rather that two different mechanisms control scaling properties across two different ranges, hereafter referred to respectively as the coarse (CS) and fine (FS) scaling ranges. 
We refer to this henceforth as \emph{biscaling}, as originally coined  in \cite{icassp2000IDC}.
Biscaling reported here is consistent with observations previously made in the literature for different traffics and for 
different analysis tools, for example \cite{hohn03,pseudoMF} for WAND traffic, \cite{Walter_MF_Allerton,Feldmann1998} for CAIDA and LBL traffics, and \cite{LOISEAU:2010:A} for Grid traffic. 

\subsubsection{Frontier scale} 

The frontier scale $\Delta_F = \Delta_0 2^{J_F}$, empirically defined as the scale connecting the CS and FS asymptotic scaling, 
is estimated as follows: 
First,  CS and FS scaling {\it best fits} are estimated independently on chosen CS and FS scaling ranges~; 
Second, departures from CS and FS {\it best fits} are computed across all scales~; 
Third, $\Delta_F $ is estimated as the first (finest scale) zero-crossing of the difference between the absolute values of these departures.

Fig.~\ref{fig:FS} shows that estimated $\Delta_F $ values remain remarkably constant along the $14$-years, with a slow and mild decrease from $0.5$s in 2001 to $0.25$s in 2014,
or equivalently that $J_F$ ranges essentially within $J_F \in (10, 13)$ corresponding to 
$ 128 \makebox{ms}  \leq \Delta_F = \Delta_0 2^{J_F} \leq 1 \makebox{s}. $ 
The 3-day trace was collected in 2013, and 
Fig.~\ref{fig:LD3Da} indicates a $\Delta_F \simeq 0.25$s, consistent with Fig.~\ref{fig:FS}.
Such orders of magnitude for $\Delta_F $ are remarkably consistent with \emph{knee position} reported in \cite{hohn03,pseudoMF,Walter_MF_Allerton,Feldmann1998,LOISEAU:2010:A}, 
though measured on different traffics and networks. 

\subsubsection{Coarse scales} 
\label{sec:CSb}

Empirical inspections of LDs, assisted with bootstrap based confidence intervals and goodness-of-fit tests (cf. Section~\ref{sec:estim} and \cite{WENDT:2007:E,Leonarduzzi2014}), for both the 15-min and 3-day traces (cf. Fig.~\ref{fig:LD3Da}) indicate the onset of CS scaling at roughly  $2 \Delta_F $. 
They also show that scaling at CS continues up to the coarsest available scale, mostly controlled by the data observation duration $\Delta_D= \Delta_0 2^{j_D}$.
The observation that scaling holds up to data duration is consistent with numerous earlier findings, for example \cite{roughanveitch07,Abry2010,LOISEAU:2010:A}.
Notably, the Median-LDs obtained from the entire 3-day trace (cf. Fig.~\ref{fig:LD3Da}) show that coarse scale scaling ranges from $0.5$s to $9h$, i.e., over $17$ octaves,  a very impressive observation, which, to the best of our knowledge, has never been reported so far on traffic collected on a commercial link.
Note that $\simeq 9h $ is the coarsest statistically significant scale available for the 3-Day ($=72$-hour) trace.
Indeed, for significance in estimation of the statistical properties, the coarsest available scale
is of the order of  $\Delta_D/2^{S}$, with $S$ empirically set to $3$ or $4$ depending on the wavelet time support and the targeted statistical confidence \cite{va01,WENDT:2007:E}. 
The CS scaling range thus corresponds to: 
$$2 \Delta_F \leq \Delta \leq 2^{-S} \Delta_D  \makebox{ or }  J_F + 1 \leq j \leq J_D - S. $$
In practice at CS, for reliable statistical estimation, guided by the statistical tools developed in \cite{veitch03,Leonarduzzi2014}, estimation of the scaling parameter ($S=4$) is conducted in ranges corresponding to $1 $s to $17$min ($[j^{CS}_1, j^{CS}_2]  = [13 , 23] $) for the 6h-blocks 
of the 3-day trace and, for  the 15-min traces, to $ 1 $s to $32$s ($[j^{CS}_1, j^{CS}_2]  = [13 , 18] $), for years 2001-2006 and to $ 0.5 $s to $32$s ($[j^{CS}_1, j^{CS}_2]  = [12 , 18] $) for years 2007-2014.

\subsubsection{Fine scales} 
\label{sec:FSb}

Empirically, the Median-LDs indicate that scaling at FS holds up to roughly  $\Delta_F/2$. 
While, obviously, scaling cannot exist at scales finer than the IAT $\tau_M$, 
Median-LDs clearly show that scaling 
holds right down to this scale.
The fine scale range is thus given by 
$$ \Delta^M_\tau \leq \Delta \leq \Delta_F/2  \makebox{ or }  J^M_\tau \leq j \leq J_F - 1 .$$


In theory, scaling analysis implies that the different scaling parameters ($H, c_1, c_2$) are measured over the same scale range: $j_1 \leq j \leq j_2$ for any given trace (or subtrace). 
However, the practical computation of wavelet leaders, 
though mandatory to assess the departure of $c_2$ from $0$, remains problematic at the finest scales 
because to compute leaders at scale $2^j$, one needs wavelet coefficients at finer scales:
As a consequence, wavelet leaders at the finest computed scales are biased (cf. \cite{WENDT:2007:E} for detailed discussions) 
as can be seen in Fig.~\ref{fig:LD3Da} (right plots). 
Thus, for simplicity and self-consistency, conservative FS ranges are selected so that all parameters are estimated over the 
same range.
Because the IAT decreases significantly along the 14 years (cf. Fig.~\ref{fig:FS}),
inspections of Sketch-LDs lead us to choose $[j^{FS}_1, j^{FS}_2]  = [7, 10] $ 
corresponding to $[16, 128]$ms for years 2001-2006, and  $[j^{FS}_1, j^{FS}_2]  = [4, 10] $ corresponding to 
$[2, 128]$ms  over 2007-2014. 

\subsubsection{Partial Conclusion 2} 
\label{sec:C2}

Internet traffic scaling properties are characterized by a biscaling regime, corresponding to scaling in two distinct scaling ranges.

\begin{figure}
\centerline{
\includegraphics[width=0.25\textwidth]{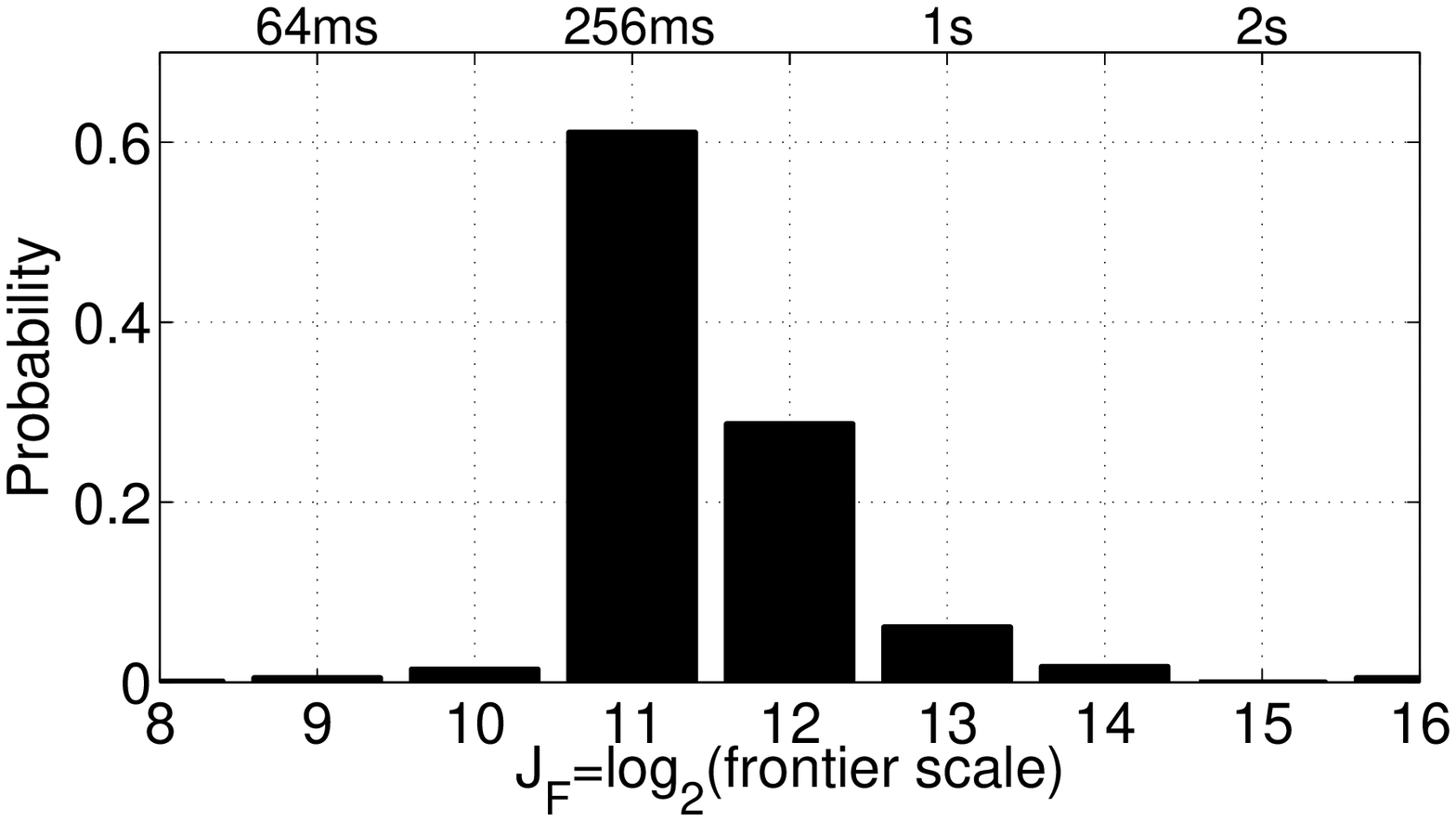}
\includegraphics[width=0.25\textwidth]{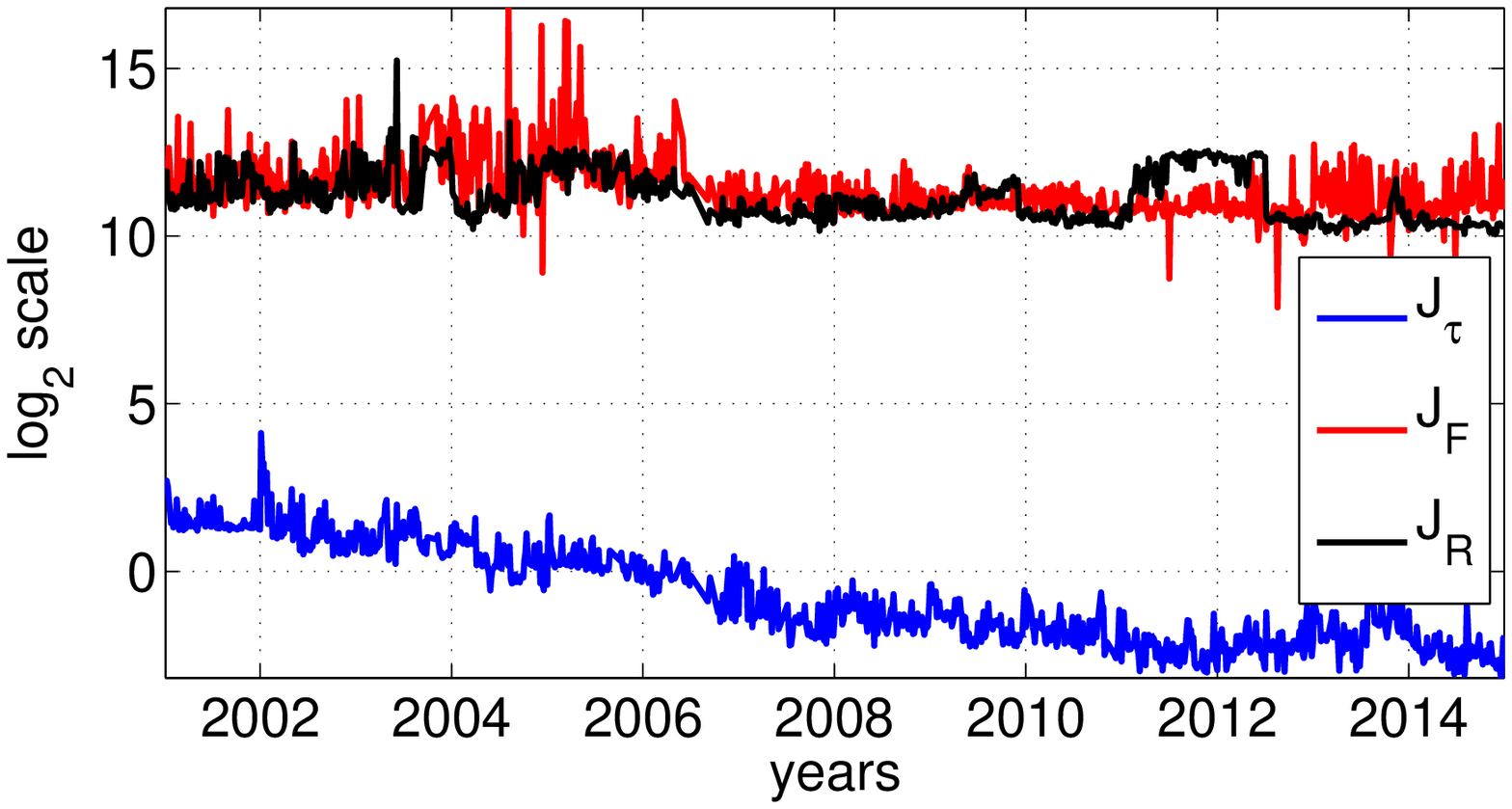}
}
\vskip-2mm
\caption{\label{fig:FS}{\bf Frontier scale.} Histogram (left) and time evolution (right) of the octave $J_F$ associated to the frontier scale.  The right plot also reports the time evolution of $J_R$ and $J_\tau$ (time scales of RTT and IAT respectively).}
\end{figure}

\begin{figure*}
\centerline{
\includegraphics[width=.33\linewidth]{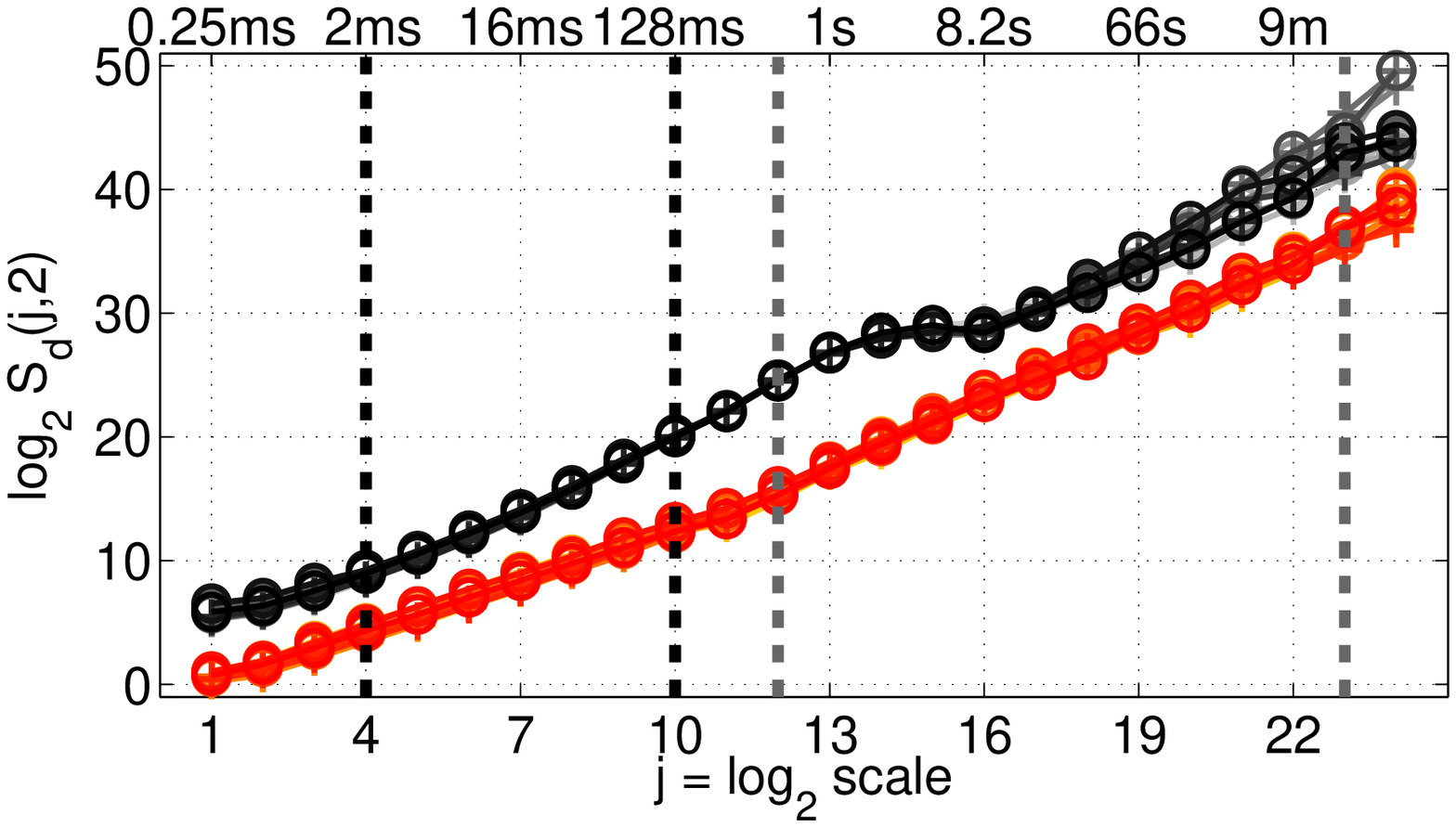}
\includegraphics[width=.33\linewidth]{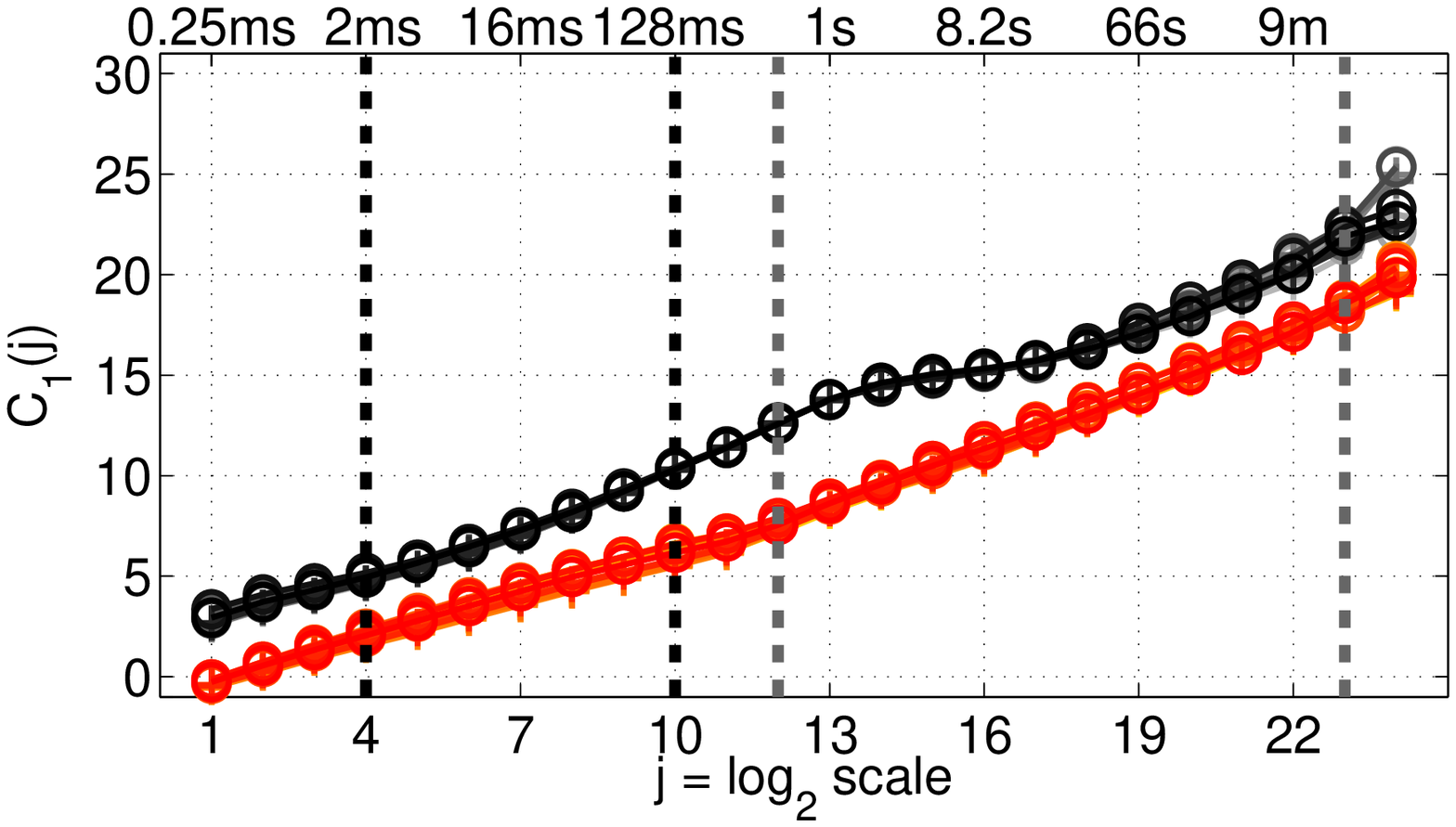}
 \includegraphics[width=.33\linewidth]{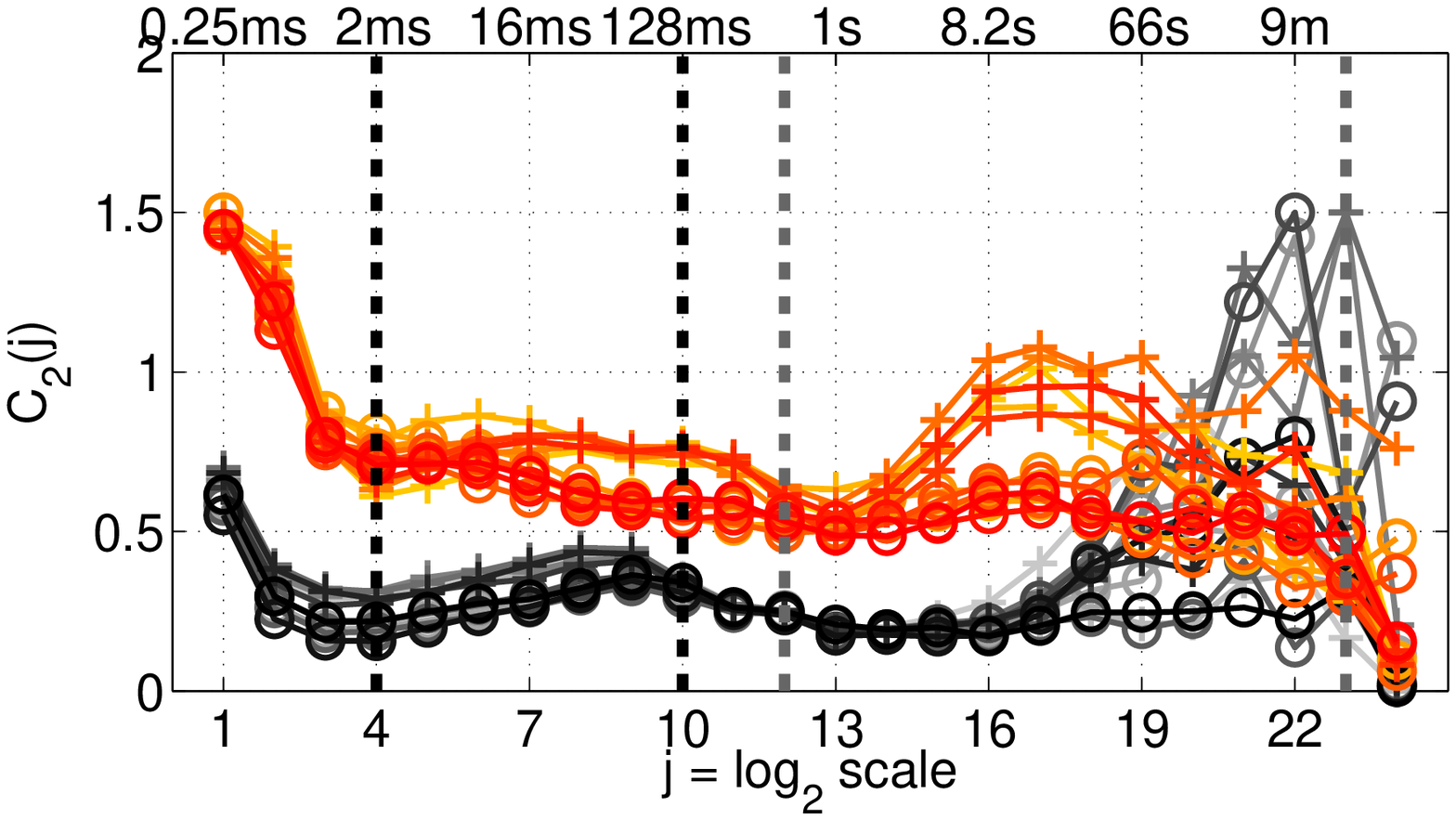}
}
\centerline{
 \includegraphics[width=.33\linewidth]{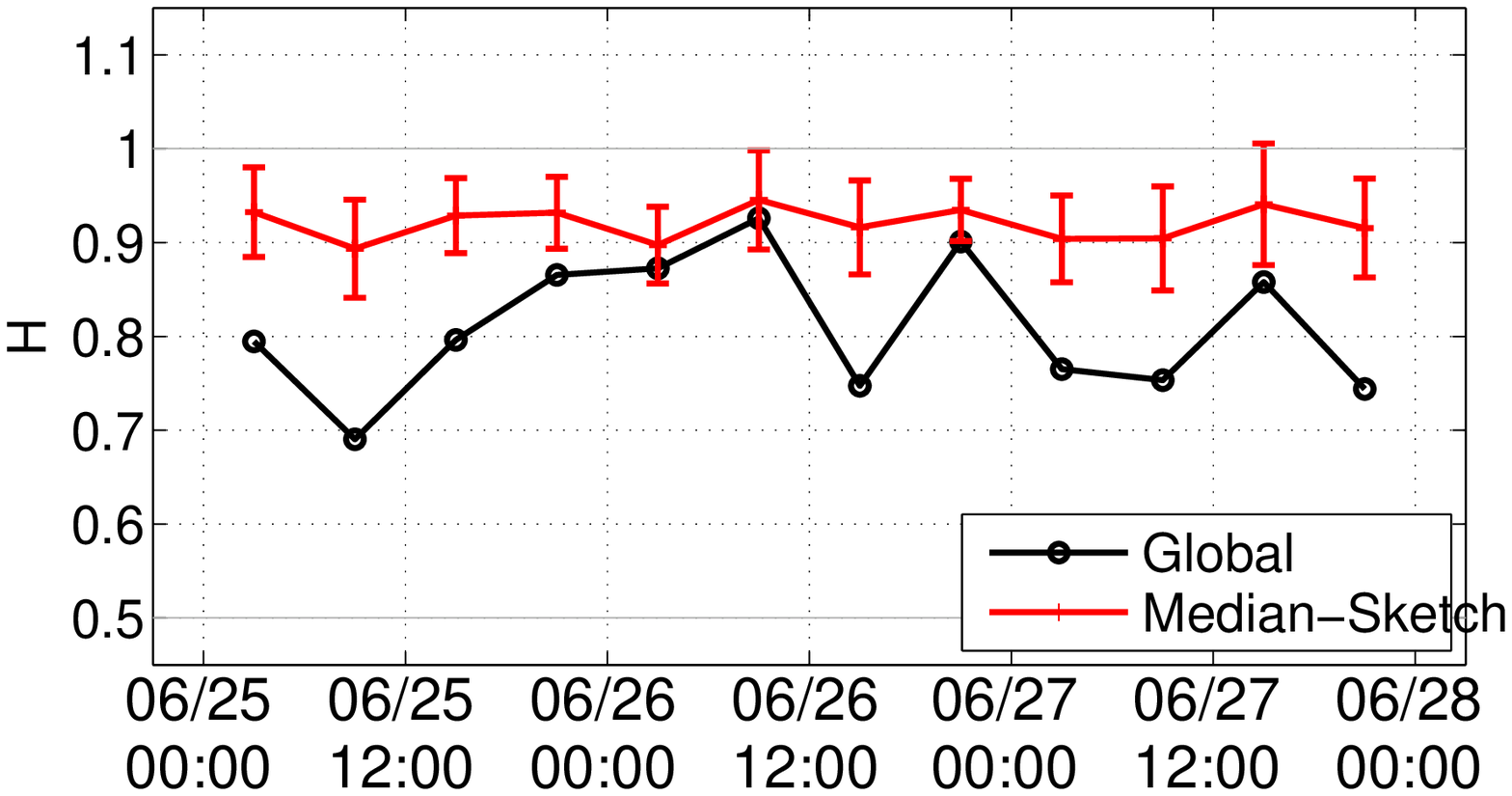}
 \includegraphics[width=.33\linewidth]{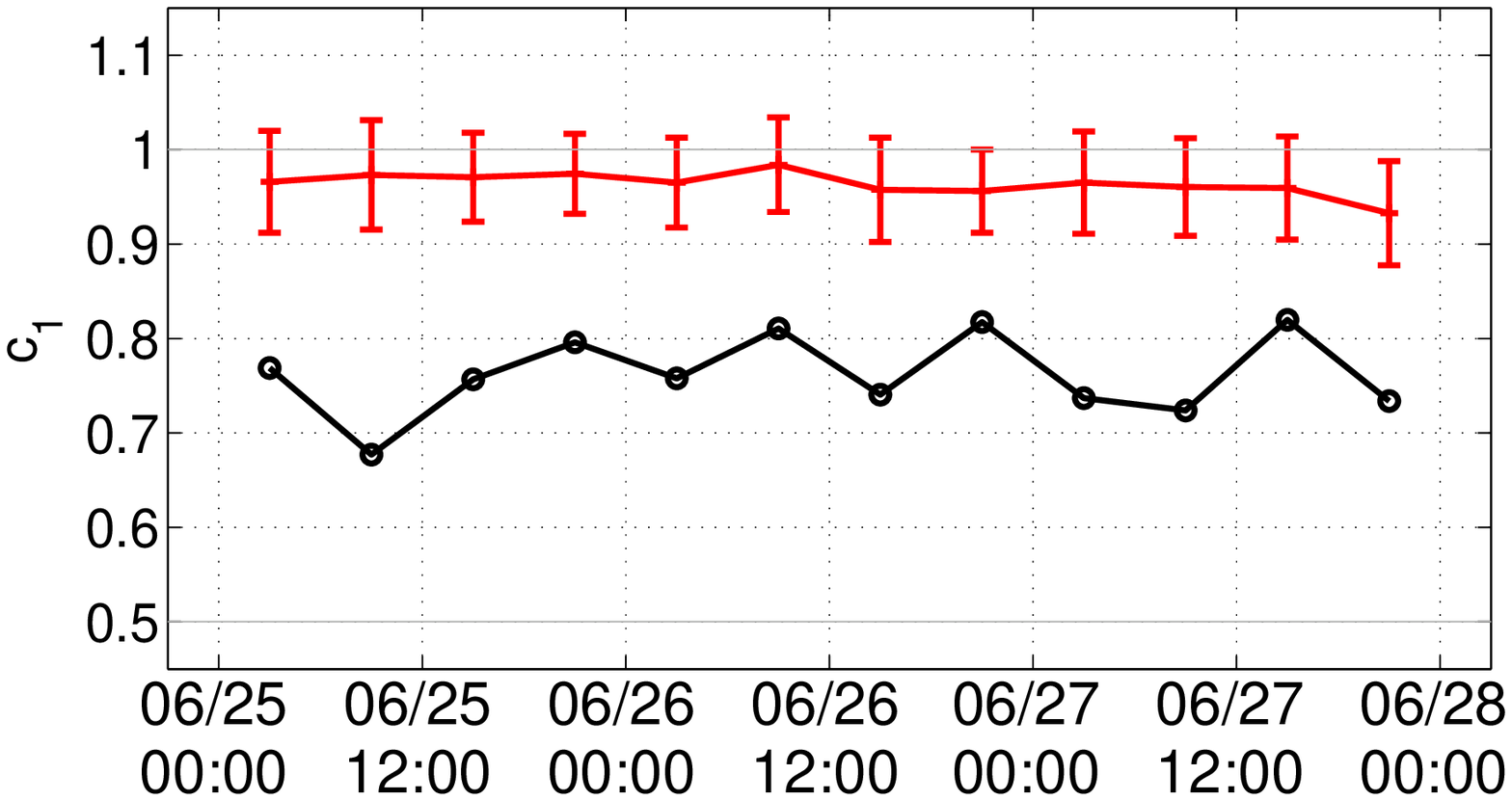}
 \includegraphics[width=.33\linewidth]{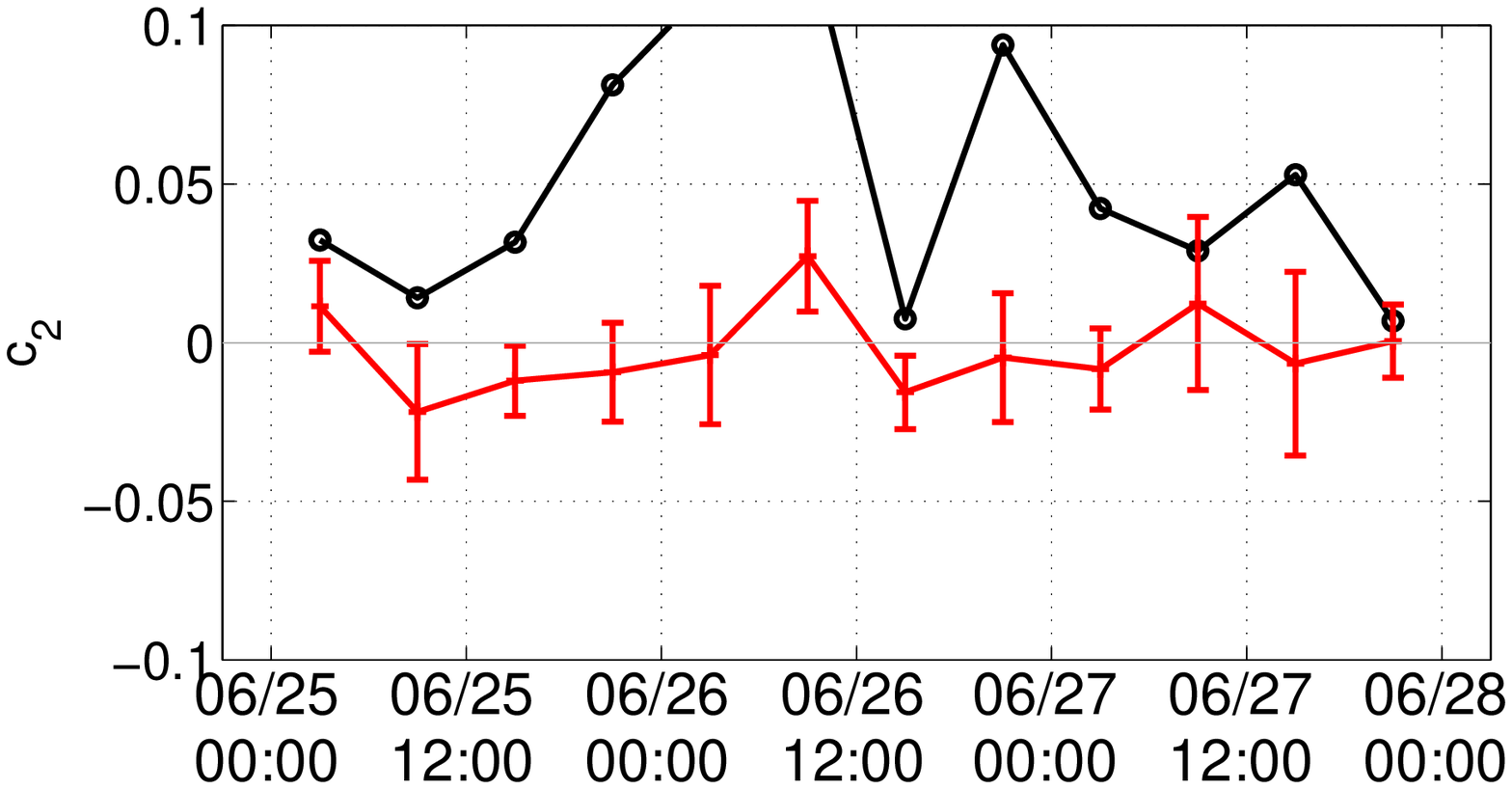}
}
\centerline{
 \includegraphics[width=.33\linewidth]{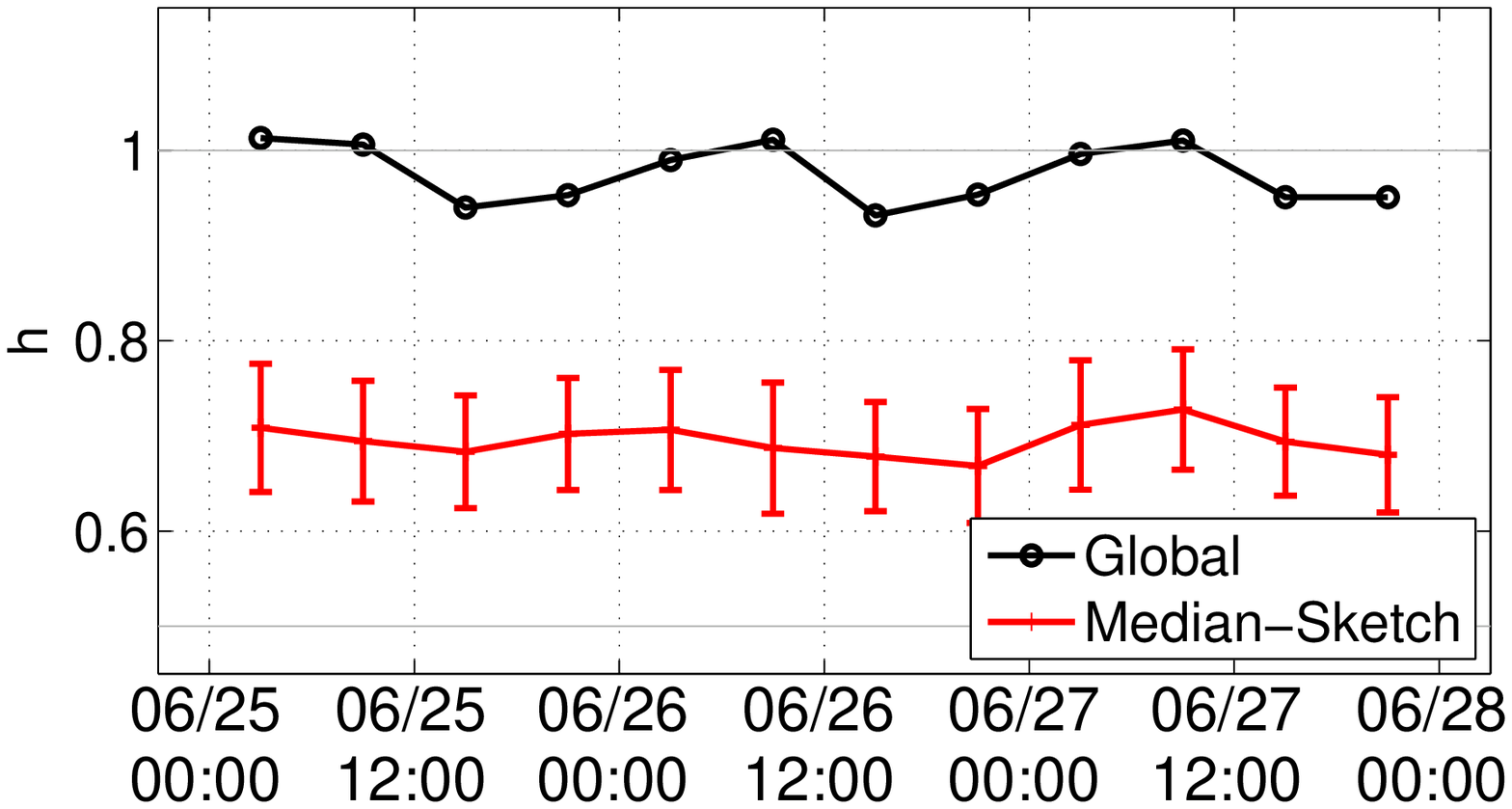}
 \includegraphics[width=.33\linewidth]{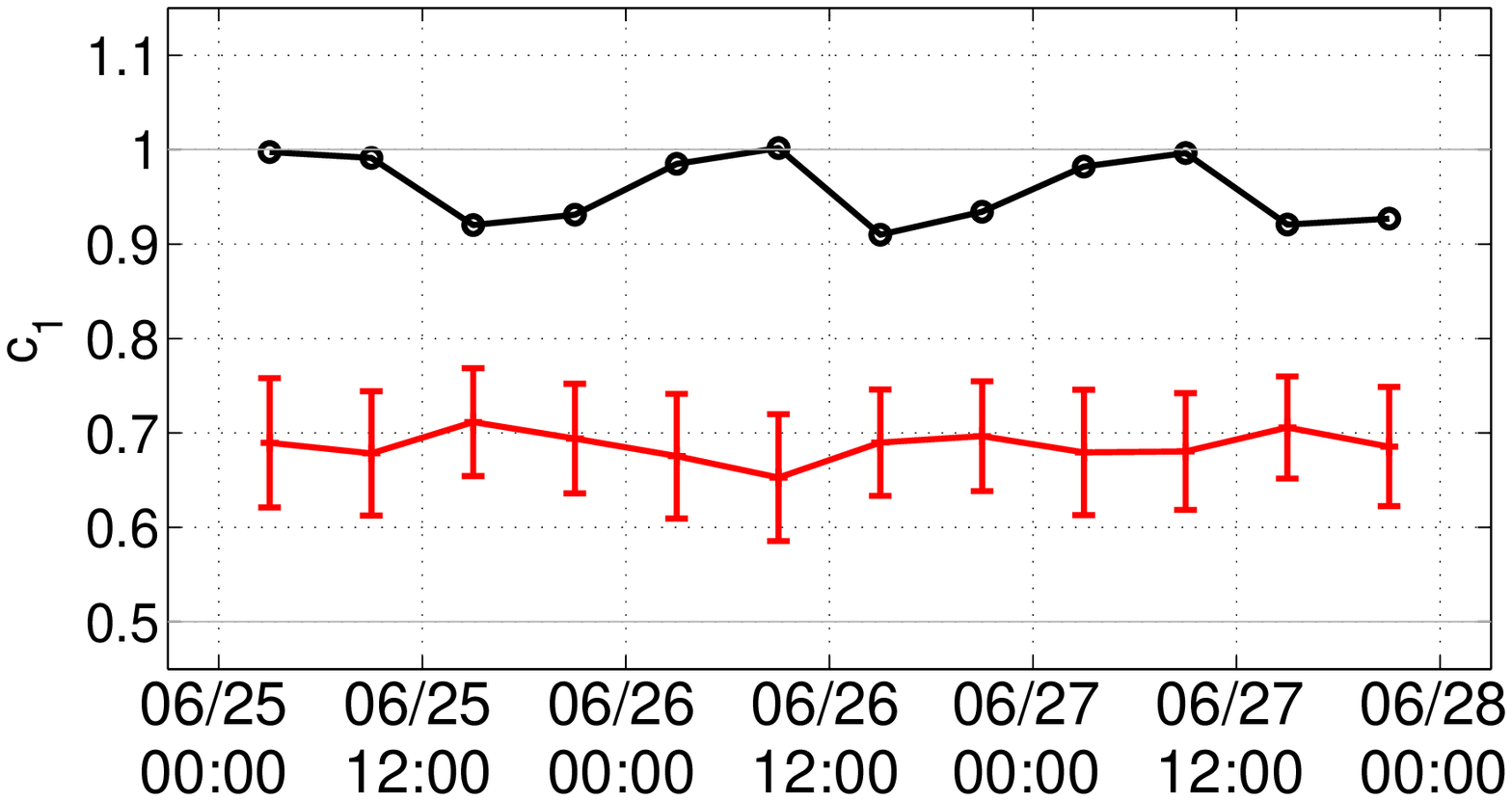}
 \includegraphics[width=.33\linewidth]{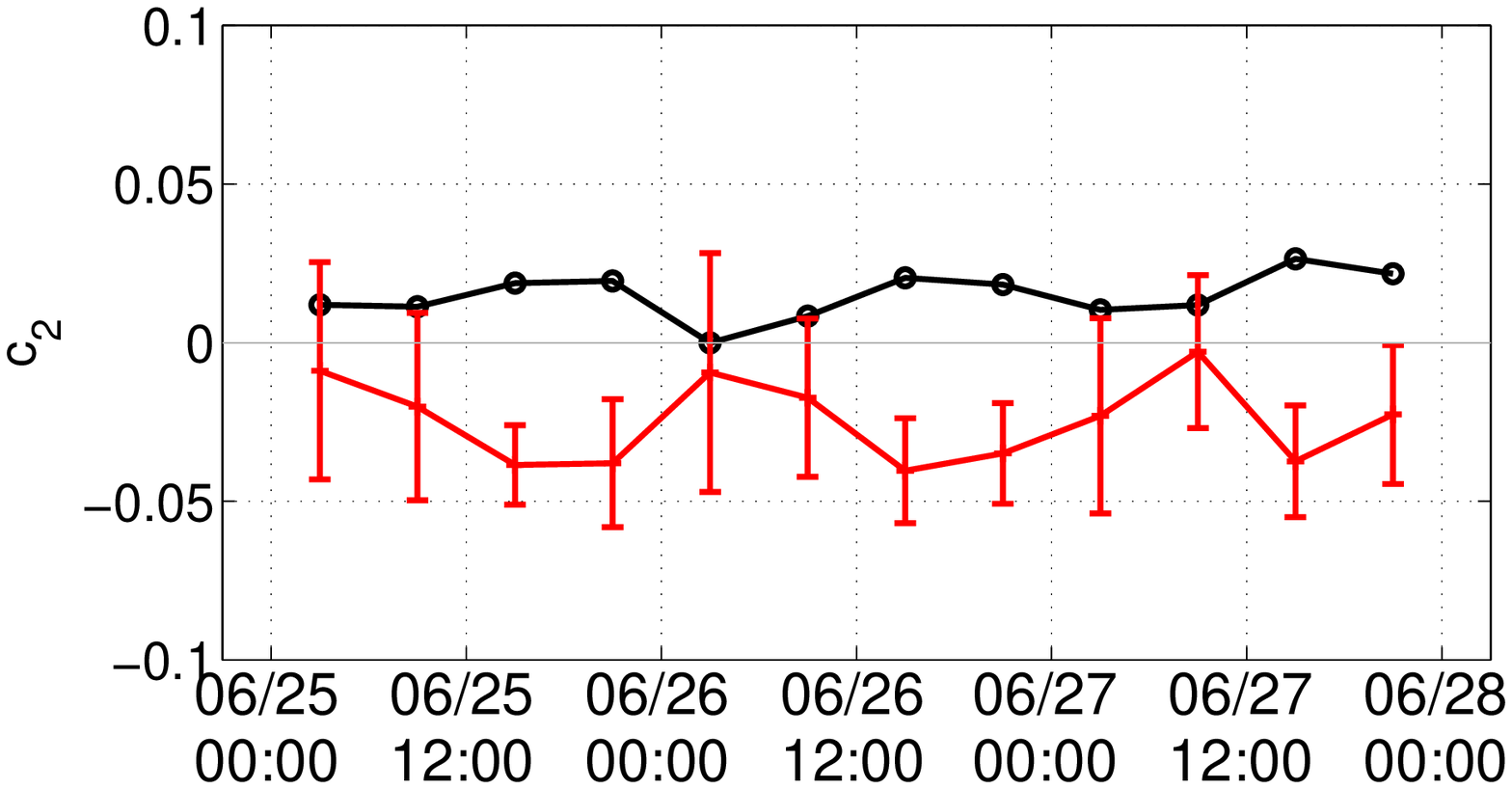}
}
\caption{\label{fig:timeevol3D}{\bf Intra-day variability along the 3-day trace.} For the 12 6h-blocks, comparisons of Global (black) and Median-Sketch (red) analyses: LDs (top); Estimated scaling exponents  at CS (middle) and FS (bottom). Confidence intervals on Median-Sketch estimates are obtained as average across sketch outputs. For LDs (top), the '+' and the 'o' correspond to night and day time estimates respectively. The dashed vertical lines mark respectively the typical IAT time scale $ J^M_\tau $ (blue), and the FS (black) and CS (dark grey) scaling ranges $(j_1,j_2)$.}
\end{figure*}

\begin{figure*}
\centerline{
\includegraphics[width=.33\linewidth]{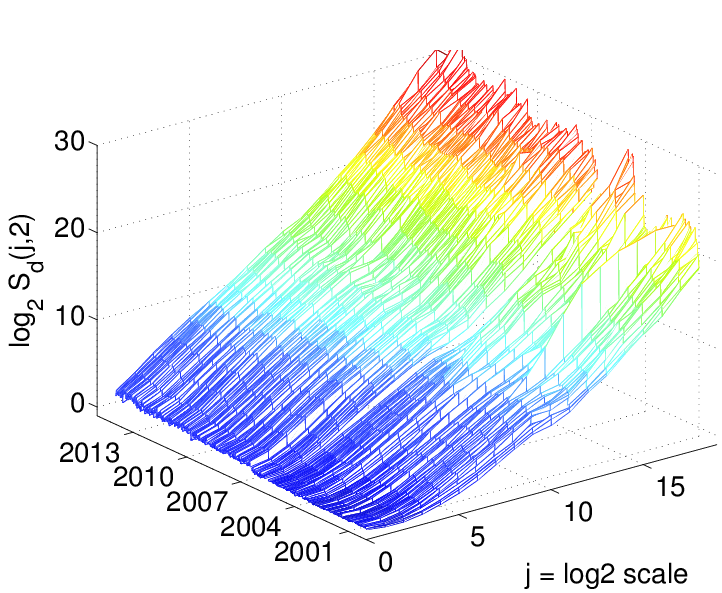}
\includegraphics[width=.33\linewidth]{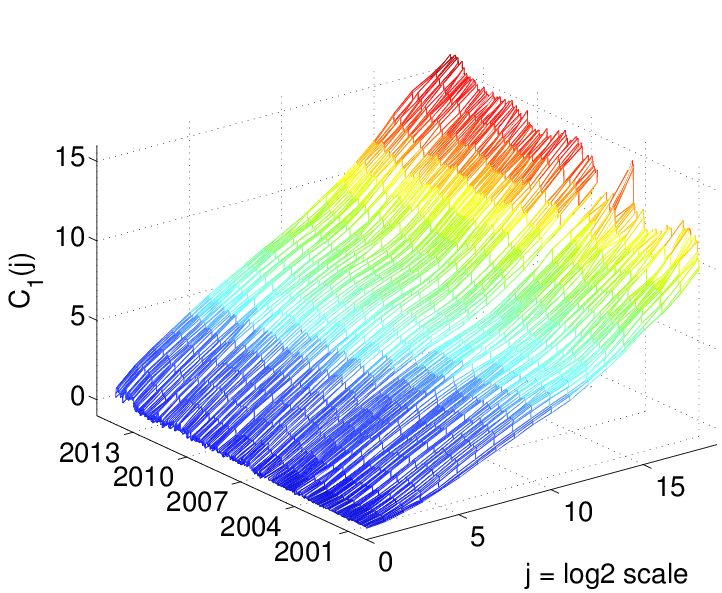}
\includegraphics[width=.33\linewidth]{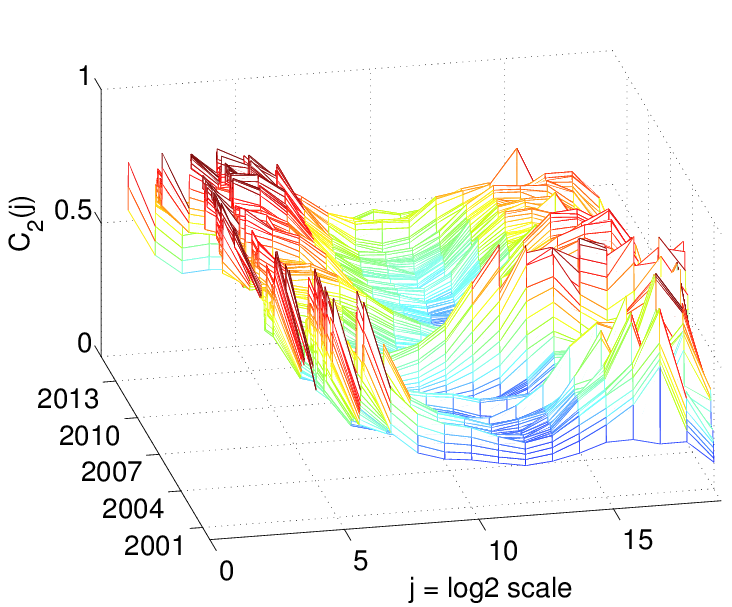}
}
\centerline{
 \includegraphics[width=.33\linewidth]{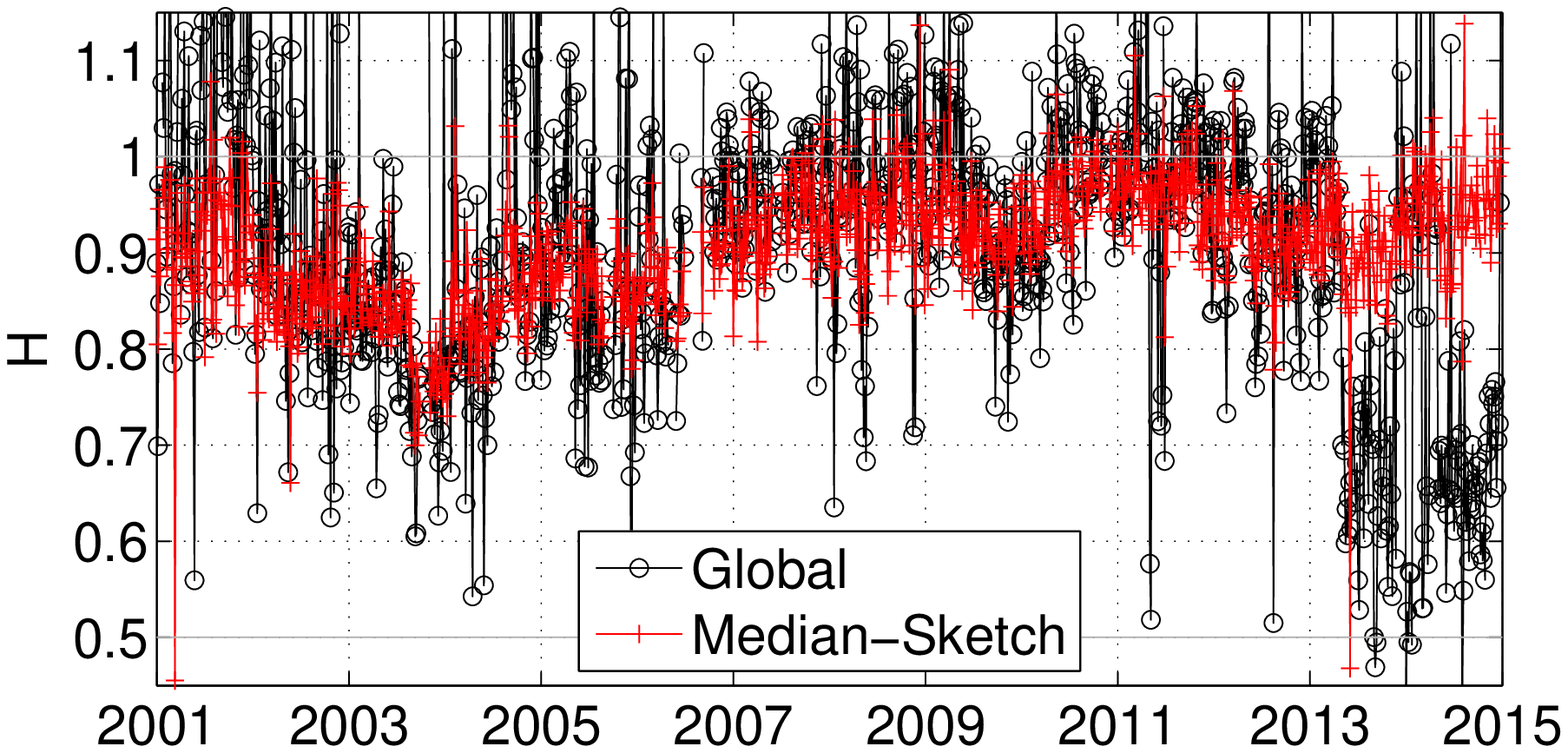}
 \includegraphics[width=.33\linewidth]{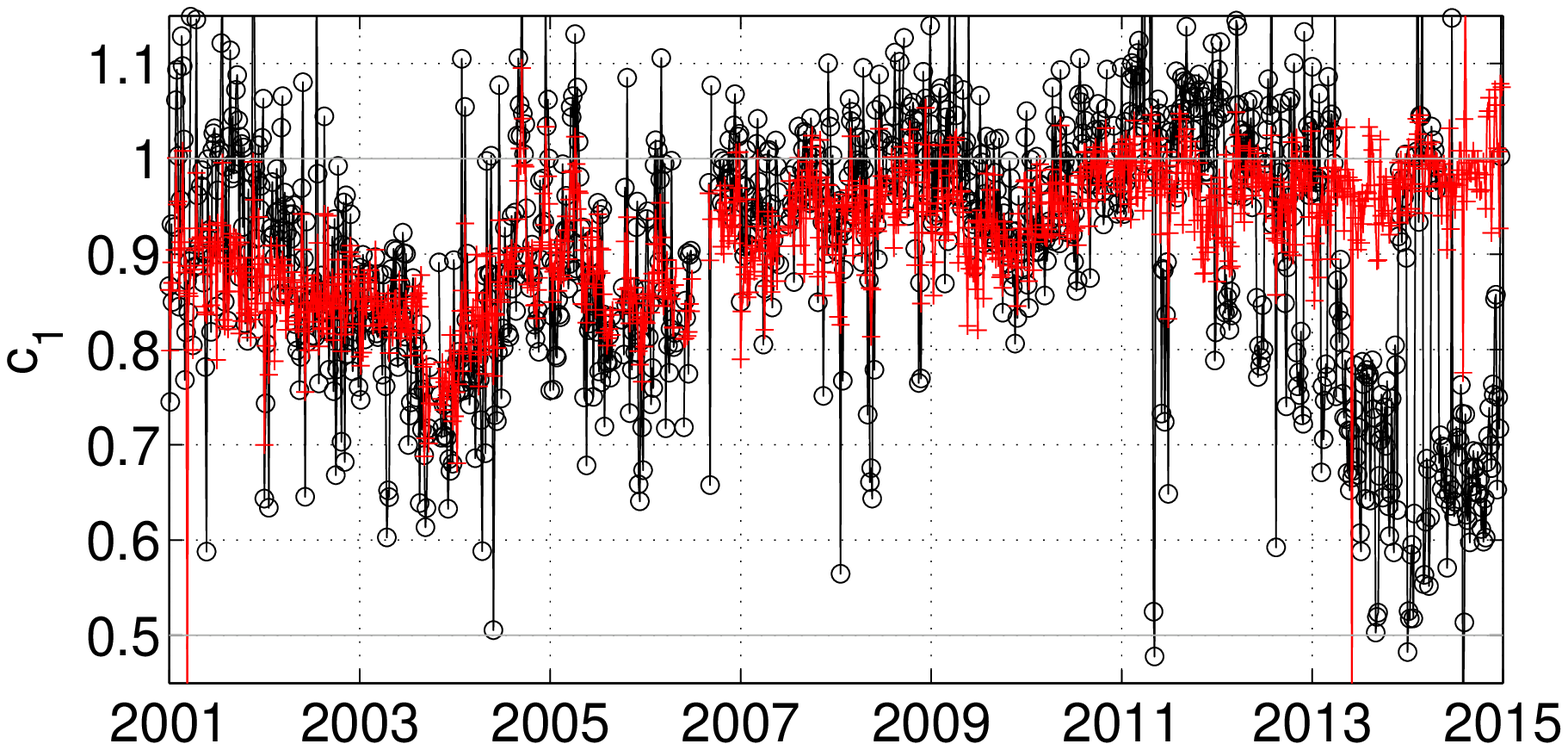}
 \includegraphics[width=.33\linewidth]{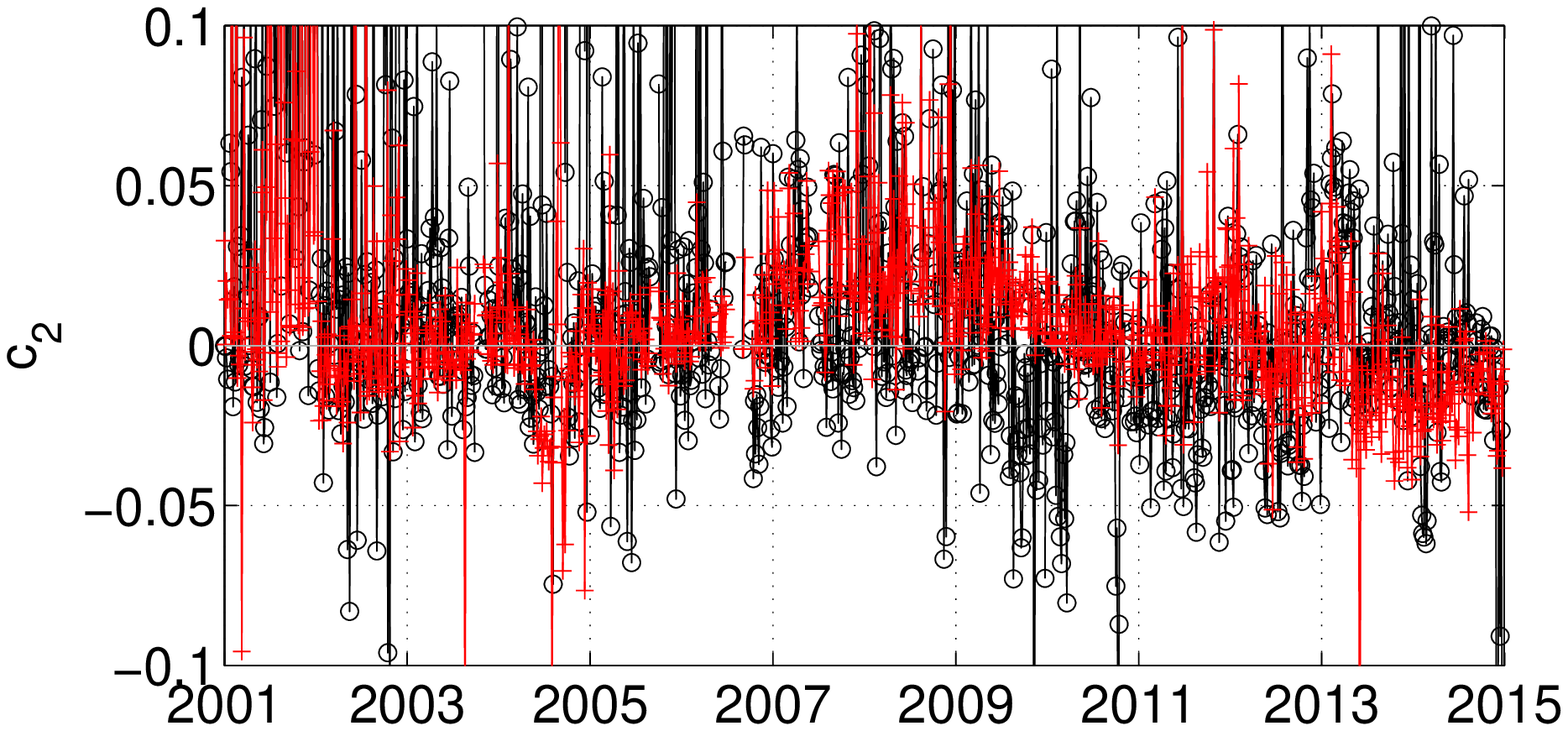}
}
\centerline{
 \includegraphics[width=.33\linewidth]{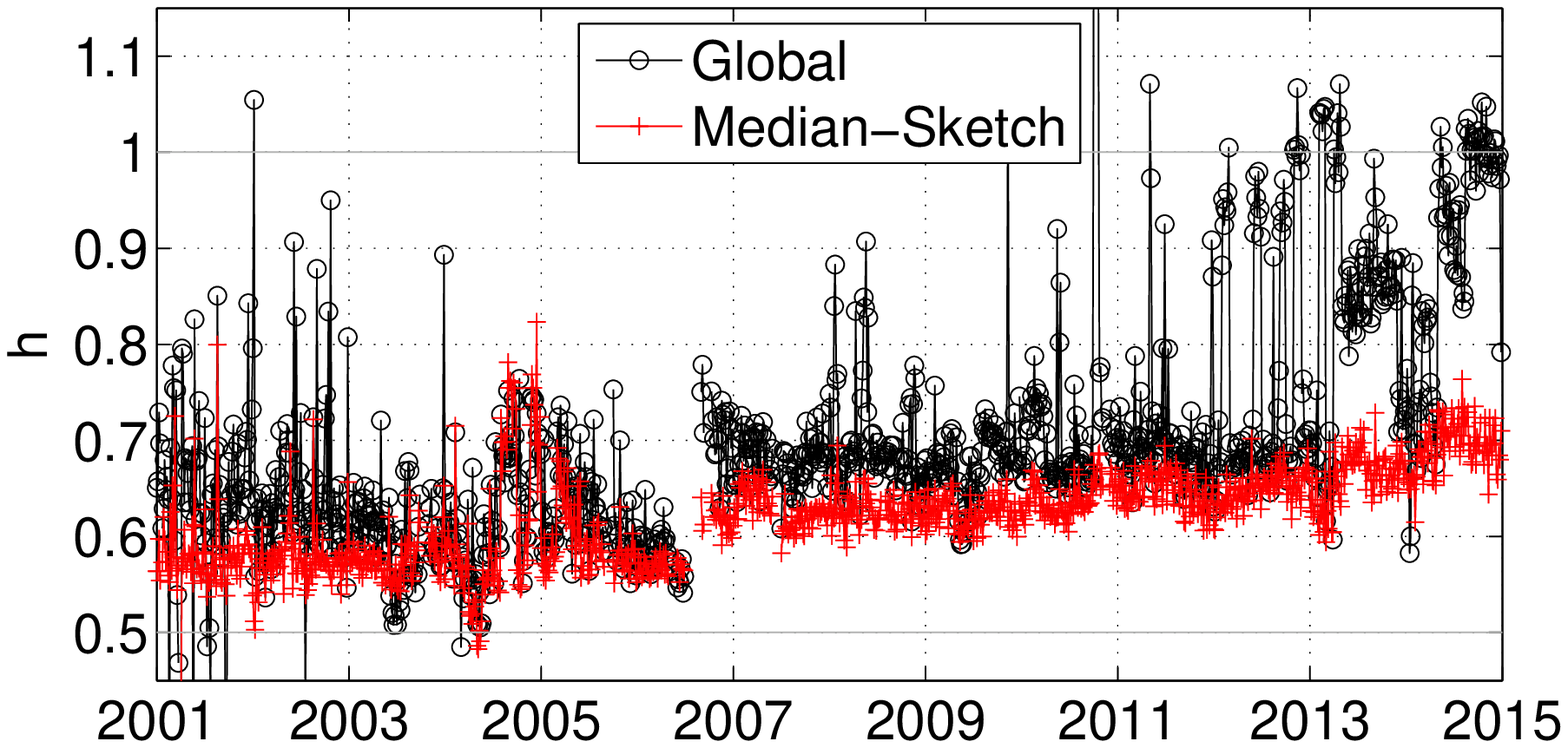}
 \includegraphics[width=.33\linewidth]{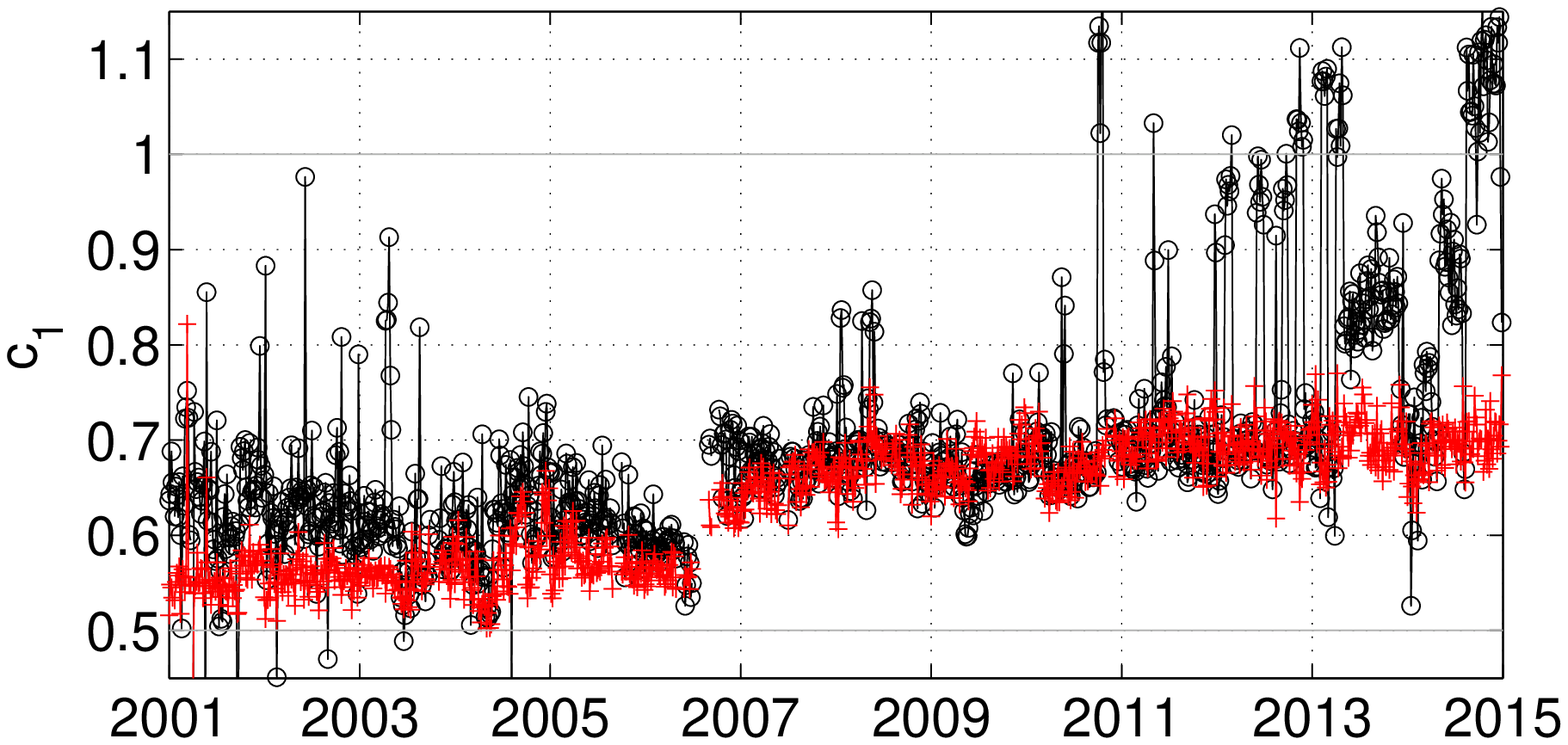}
 \includegraphics[width=.33\linewidth]{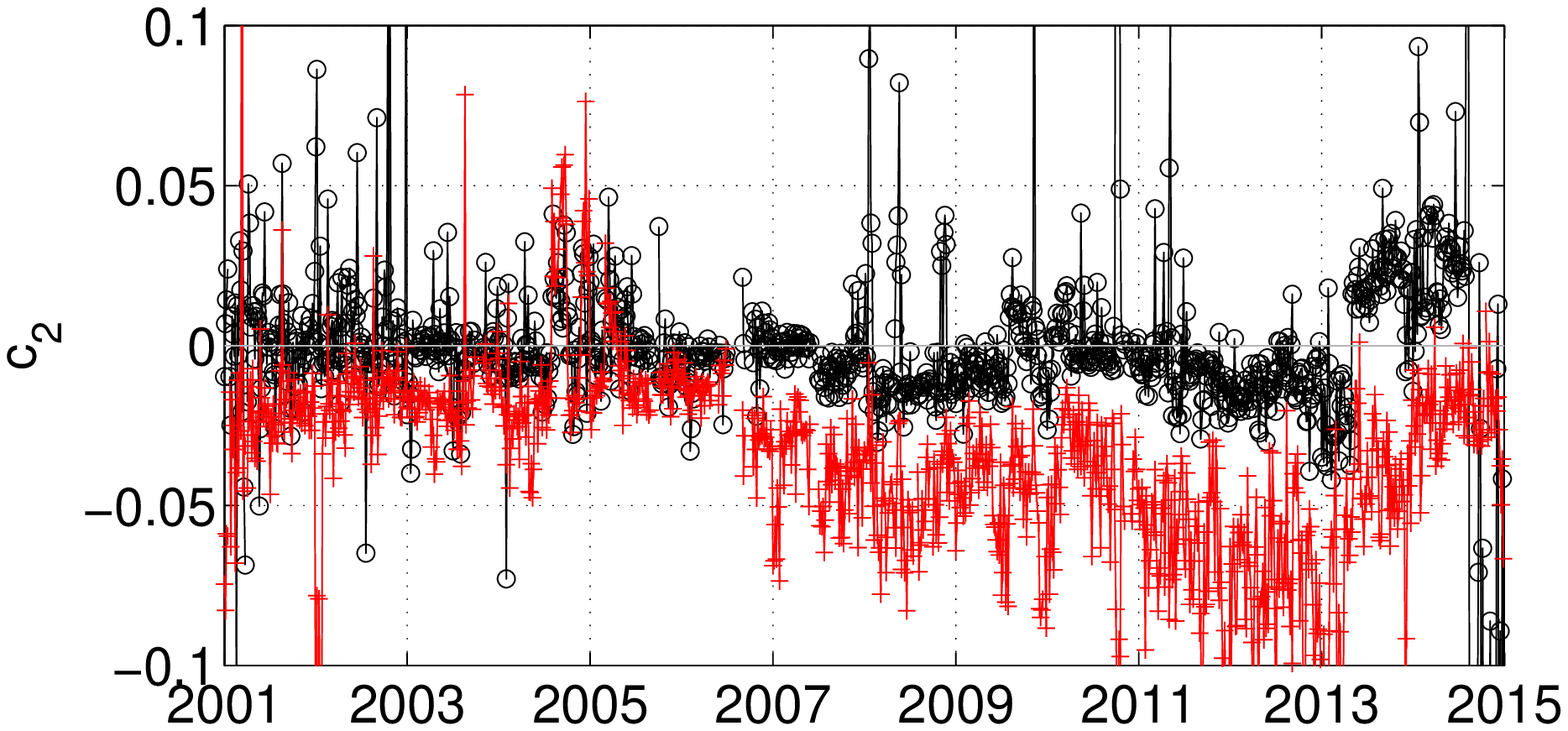}
}
\vskip-2mm
\caption{\label{fig:timeevol15min}{\bf Long term Evolution along the 14 years.} For  the 1176 15-min traces,
Median-LDs (top) and comparisons between Global (black) versus Median (red)-based estimated scaling exponents  at CS (middle) and FS (bottom).}
\end{figure*}


\section{Variability and evolution of scaling with time}
\label{sec:evol}

\subsection{3-day trace and intra-day variability} 

Fig.~\ref{fig:timeevol3D} compares Global versus Median-LDs (and corresponding scaling exponents) estimated from the 12 non overlapping 6h-traces within the 3-day trace (top row).
Global-LDs show a much larger variability, notably at CS, such that the use of Global-$C_2(j)$ becomes meaningless.
In fact, the scaling exponent estimates obtained from Global-LDs display a variability which is far too large for consistency with (bootstrap based) confidence intervals. 
Notably, the Global-LD estimates show a 24-hour periodicity, particularly clear at FS, that has (almost) disappeared from the Median-LD estimates. 
Beyond the potential \emph{natural} diurnal cycle that could explain such modulation, inspection of traffic trace showed that the \emph{trinocular} experiment, discussed earlier in Section~\ref{sec:data}, was active during this trace.
This anomalous traffic can be regarded as non stationary as it was essentially run at night (Japanese time), hence explaining the 24-hour periodicity. 
Further, the \emph{trinocular} experiment sends, every  11 minutes, 1 to 15 ICMP probes to 3.4M blocks of IP addresses. 
Probes sent to the same block are spaced out by a 3 second timeout, thus producing a specific time scale in temporal dynamics that materializes as the bump at $j \simeq 14 $ or $15$ in Global-LDs (cf. Figs.~\ref{fig:LD3Da} and \ref{fig:timeevol3D}).
On average, Trinocular sends about 19.2 probes per hour per IPv4 block. 
It 
therefore 
produces a massive ICMP packet traffic superimposed to the remainder of the regular background traffic, thus significantly affecting traffic statistics and scaling properties at all scales.
Global-LDs are hence polluted by this anomalous traffic, both at CS and FS, and their use would lead one to conclude that traffic undergoes a periodic modulation of its statistics and scaling properties, while this is actually due to the intermittent occurrence of the anomaly.
Conversely, Median-LDs  provide practitioners with a robust characterization of the background traffic, not altered by the \emph{trinocular} anomaly.
Median-LDs (and corresponding estimated parameters) display a remarkable constancy over time along the 3 days, thus showing the stationarity of intra-day statistical properties of Internet traffic, with minimal impact of the diurnal cycle.
 
Interestingly, a careful inspection of the Median-LDs for $C_2(j)$, Fig.~\ref{fig:timeevol3D} (top right), still shows a residual 24-h modulation ($C_2(j)$ computed during 6-h day-time blocks differ from those computed during 6-h night-time blocks). 
The Source IP Address has been chosen here as flow (hashing) key for sketching traffic. 
This allows \emph{trinocular} (produced from a same IP Address) traffic to be concentrated into a single sketch output. 
However, this generates a response traffic, far lower in volume yet anomalous, which is not similarly concentrated.
Robustness to that response traffic is indeed not achieved by hashing on Source IP Address, but would instead be obtained using Destination IP Address as the hashing key. 
In practice, one should thus ideally perform hashing on both Source and Destination Addresses. 
This indicates that LD $C_2(j)$ corresponding to a refined and detailed analysis of statistical properties at all statistical orders, i.e., beyond correlation, may thus be more impacted by remaining anomalous traffic than are LDs $C_1(j)$ and $\log_2 S_d(j)$, which essentially quantifies the 2nd order statistics.  
\vspace{-2mm}

\subsection{15-min traces} 
\vspace{-1mm}

Fig.~\ref{fig:timeevol15min} reports, for the 15-min traces,  the Median-LDs (top) and compares the scaling exponents, as a function of trace collection time, estimated from Median-LDs to those of Global-LDs, for CS (middle) and FS (bottom).
Global-LD estimates show a very large daily variability, far too large to be consistent with statistical estimation fluctuations. 
Common practice would trust such estimates, and lead to the (incorrect) conclusion that traffic scaling is not a robust property, as estimates keep changing.
However, automated inspection of MAWI traces shows that there is almost no single day without significant anomalies \cite{Dewaele2007,borgnat:infocom2009,romain:conext2010,mazel:trac14}.  
Global-LDs are thus essentially shaped by anomalous traffic.
Conversely, Sketch-LDs (top row) and the corresponding estimated scaling parameters display a significantly reduced variability from one day to the next.
Such variability is consistent with bootstrap-estimated statistical fluctuations, following procedures well-assessed in \cite{WENDT:2007:E}.
These observations constitute a significant indication for constancy of CS scaling over the 14 years.

Further, 
inspection of Fig.~\ref{fig:timeevol15min} shows an actual (mild yet clear) change in scaling exponents and scaling ranges, separating two roughly piecewise constant periods, from Jan. 2001 to June 2006 and from Oct. 2006 to Dec. 2014, respectively. 
Interestingly, summer 2006 corresponds to the link update, mentioned in Section~\ref{sec:data}.
This shows that a network reconfiguration, even if major (significant increase of the available bandwidth),
does not change drastically the general shape of the scaling properties in traffic (notably biscaling remains), 
but affects, though
 only marginally, scaling exponents and scaling ranges:
Notably, $\Delta_F$ and $\Delta_\tau$ are both slightly decreased (cf. Fig.~\ref{fig:FS}), which motivates the changes in the scaling range selection reported in Sections~\ref{sec:CSb} and \ref{sec:FSb}. 
The remaining variations of Median-LD estimates of $H $ at both CS and FS, around years 2004-05 (cf. Fig.~\ref{fig:timeevol15min}), correspond to the period of intense  Sasser virus traffic. 
They show that once a given anomalous behavior becomes the dominant traffic, the sketch procedure considers it as the \emph{normal} traffic and ceases to provide robustness against it \cite{borgnat:infocom2009}. 
\vspace{-3mm}

\subsection{Partial Conclusion 3} 
\vspace{-1mm}

Median-LDs show unambiguously that Internet traffic exhibits remarkable constancy of its statistical and scaling properties, 
both at CS and FS, both for intra-day variability, with no impact 
of diurnal cycles acting as a nonstationary trend, and for long-term evolution: 
The scaling  properties in MAWI traffic do not significantly change along the 14 years, neither in the shape of the LDs (biscaling is a robust property) nor even in the value taken by the scaling exponents, and this, despite the major changes undergone by the Internet during the last decade.

\section{Natures and origins of scaling}
\label{sec:nature}


This section investigates the nature of scaling (LRD or multifractality), by systematically analysing scaling parameters $H, c_1$ and $c_2$ estimated off Median-LDs, at both CS and FS. 
Potential mechanisms for the origins are also investigated quantitatively.
\vspace{-2mm}

\subsection{Coarse scales}
\label{sec:CS}

\subsubsection{Nature of Scaling}

For the 6-hour blocks of the 3-day trace, Median-LDs estimate $H \simeq 0.92 \pm 0.03$, consistently along the 3 days (Fig.~\ref{fig:timeevol3D}, middle left plot). 
This is consistent with the 15-min traces where $H$ is observed to remain confined in the range $ 0.8 \leq H \leq 1$, with typical values around $H \simeq 0.94 \pm 0.03 $ after 2006, while $H$ is found slightly lower $H \simeq 0.86 \pm 0.04$ before 2006 (Fig.~\ref{fig:timeevol15min}, middle left plot).
Such values of $H$ are extremely consistent with earlier measures on the same traffic \cite{borgnat:infocom2009} as well as on many other different traffics \cite{Leland1994,AbryVeitch98,va99,hohn03,pseudoMF}. 
It is also consistently 
observed that  $c_1 \simeq H$ and $c_2 \simeq 0$ (see Figs.~\ref{fig:timeevol3D} and \ref{fig:timeevol15min}, middle row, left plots) thus indicating no multifractality at CS.
This is consistent with \cite{pseudoMF} that reported no evidence of multifractality on Auckland Traffic in an equivalent range of time scales.
It is also interesting to note that besides the decrease of $\Delta_F$ (visible in Fig.~\ref{fig:FS}, left),  the link upgrade in 2006 has a fairly limited impact on scaling at CS.

\begin{figure}[h]
\centerline{ \includegraphics[width=.66\linewidth]{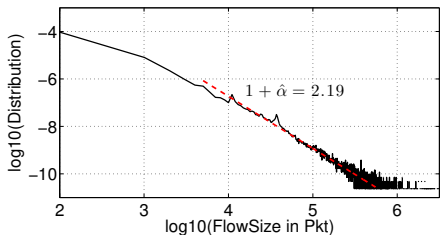}}
\vskip-2mm
\caption{ \label{fig:Halphab} {\bf Origin of scaling at CS: LRD versus heavy Tail.} Flow size distribution (measured in pkt)
 show heavy-tail behavior, with tail exponent in exceptional quantitive agreement with Hurst exponent $H$ at CS, as predicted in \cite{onoffproof}.
 }
\end{figure}

\subsubsection{Origins}

As recalled in Section~\ref{sec.motiv}, a mechanism was proposed to explain LRD in Internet Traffic, by relating it to 
the distribution of Internet object sizes via a generic heavy tail On/Off superimposition mechanism \cite{Leland1994,onoffproof}. 
In essence, this theoretical mechanism predicts asymptotic scaling in the limit of very large scales: 
it is hence naturally associated to scaling at CS as defined in the present study.
It, however, raises the issue of selecting the range of scales where LRD should be practically observed, in relation to heavy tails.
As reported in Section~\ref{sec:CS}, CS scaling extends up to a scale corresponding to data recording duration, and down to 
cut-off scale $2^{J_F}$. 
These findings, associated with the typical $H \simeq 0.9$, are in agreement with the methodological analyses of this asymptotic mechanism reported in \cite{roughanveitch07,Abry2010,LOISEAU:2010:A}.
Further, the theorem in \cite{Leland1994,onoffproof} predicts that, in the limit of CS, Internet traffic time series should be asymptotically Gaussian self-similar, thus excluding multifractality (i.e., $c_2 \equiv 0$), again in agreement with empirical measurement reported in the present study. 

Despite the numerous efforts reported in the literature (cf. e.g., \cite{Crovella1997,Willinger2004}), a quantitative validation of the theoretical relation between heavy tail index $\alpha$ and $H$ has turned out to be difficult to obtain from real measurements.
This can been explained by practical difficulties in measuring the actual Internet object distribution and its corresponding index, as thoroughly documented in  \cite{Willinger2004,roughanveitch07,Abry2010,LOISEAU:2010:A}. 
Following insights offered by the use of the Cluster Point process (CPP) model in \cite{hohn03}, we explore here this generic link between LRD and heavy tail by studying the distribution of the flow sizes (in number of pkts). 
The estimation procedure for $H$ and $\alpha$,  in particular in terms of selecting the ranges of scales and quantiles over which to conduct the linear regressions, carefully follows the methodology devised in \cite{Abry2010,LOISEAU:2010:A}.
The combined use of two methodological ingredients (multiscale analysis and random projections) with the
 exceptional duration of Internet data (3-day trace), enables us to measure, on one hand $ H = 0.9\pm 0.05 $ for the pkt arrival process, and on other hand, $ \alpha = 1.19 \pm 0.05$ as the tail index of the flow size distribution, as illustrated in Fig.~\ref{fig:Halphab}. 
The theorem in \cite{Leland1994,onoffproof} predicts a relation $H = (3- \alpha) / 2$, which turns here into a remarkable match $ (3 - 1.19)/2 = 0.905 \pm 0.025$.
To the best of our knowledge, this constitutes a quantitative agreement of unprecedented-quality between the theoretical prediction and empirical measures obtained on actual Internet traffic collected on real commercial links (see a contrario \cite{Abry2010,LOISEAU:2010:A} for simulated or Grid traffics).   

\subsubsection{Partial conclusion 4}

The present longitudinal study clearly shows robust and strong LRD with no multifractality for Internet traffic at CS, i.e., beyond $1$s, moreover in close quantitative agreement with the tail exponent of flow size.

\begin{figure*}
        \subfloat[Top, empirical histogram of RTT, class-partitioning. Bottom, Median-LDs conditioned to RTT quantiles. 
The dashed-vertical lines correspond to the median RTT for each class of flows. LDs are normalized in amplitude for comparison.\vspace{3mm}]{
	      \begin{minipage}{0.3\textwidth}
                        \includegraphics[width=\textwidth]{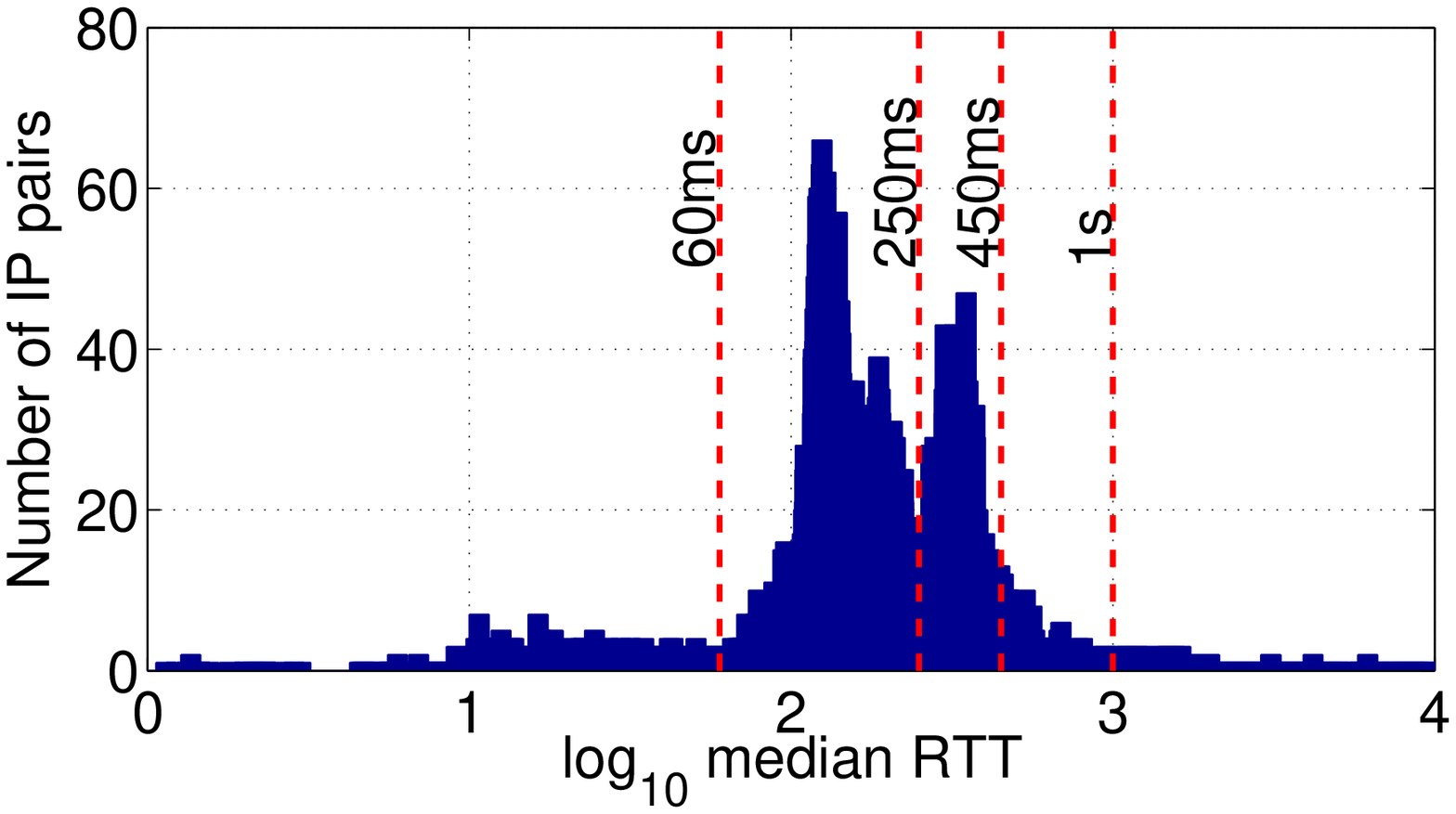}
                \includegraphics[width=\textwidth]{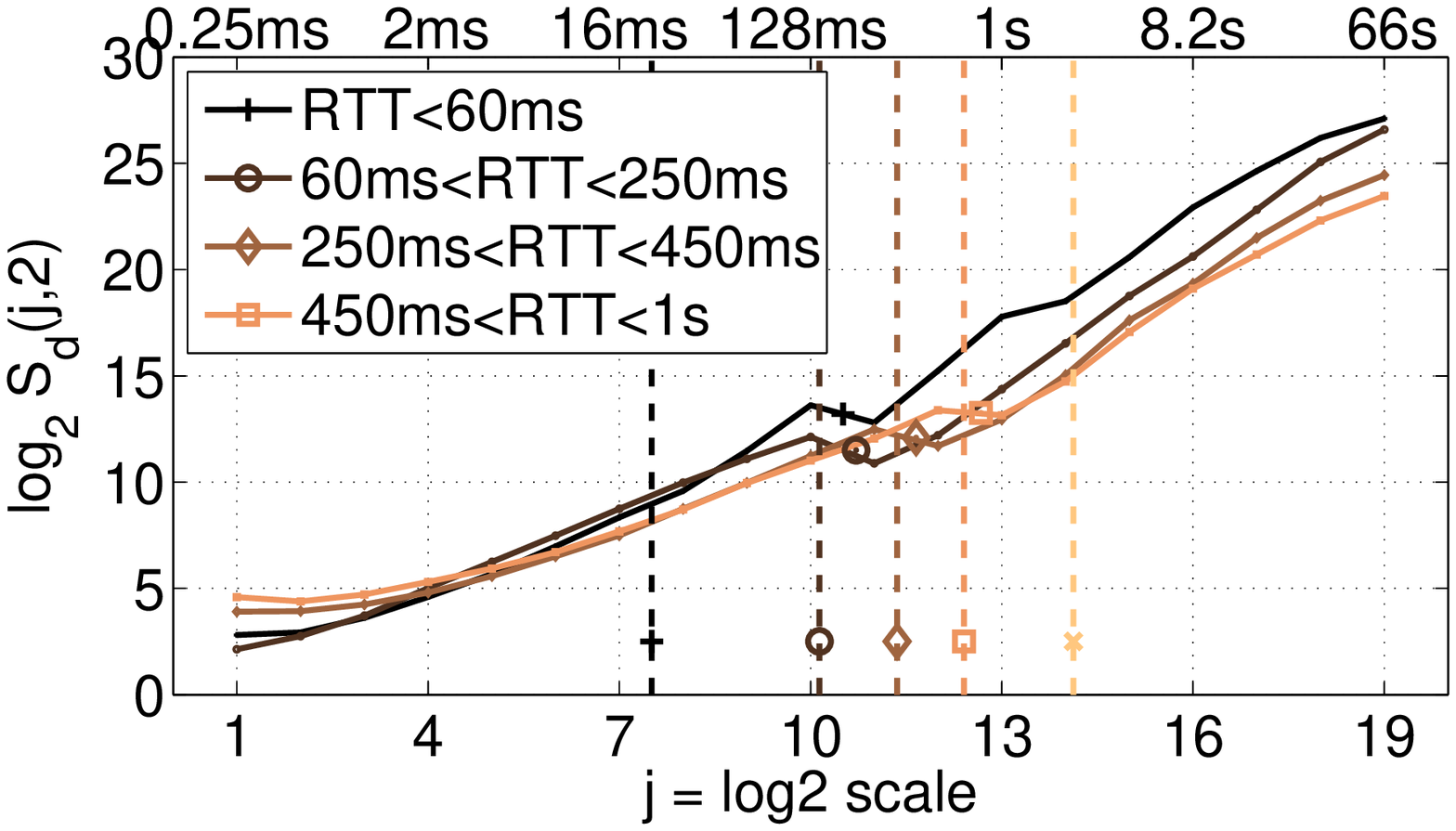}
                \label{fig:LRDRTTa}
                \end{minipage}
        }
        \,
        \subfloat[Top: $J_F$ function of $J_R$. Bottom, $c_1$ function of $J_M$. The markers and colors of the data indicate the median RTT, as per the figure (a) bottom, on the left.\vspace{5mm}]{
        \begin{minipage}{0.3\textwidth}
                       \includegraphics[width=\textwidth]{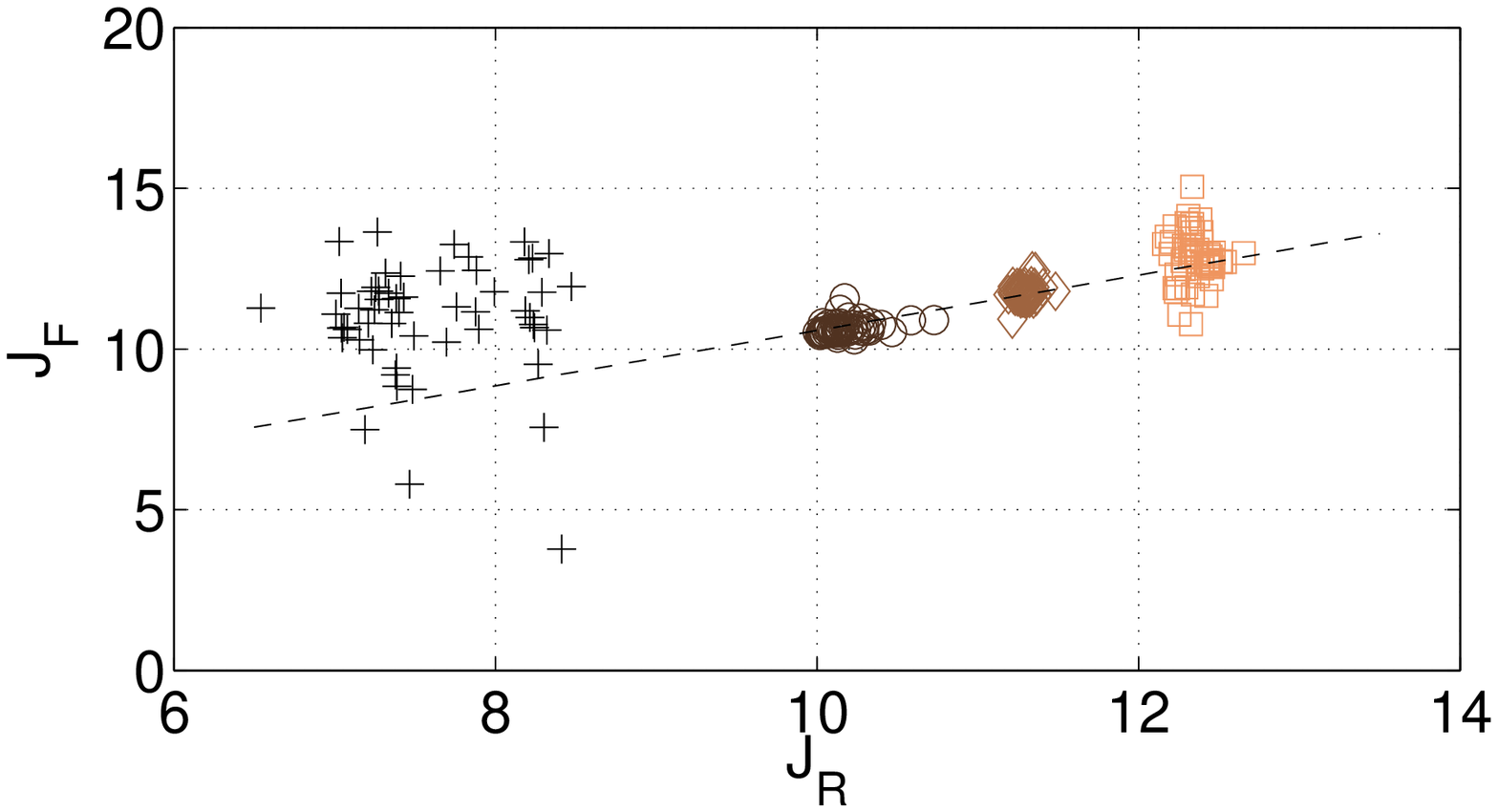}
                \includegraphics[width=\textwidth]{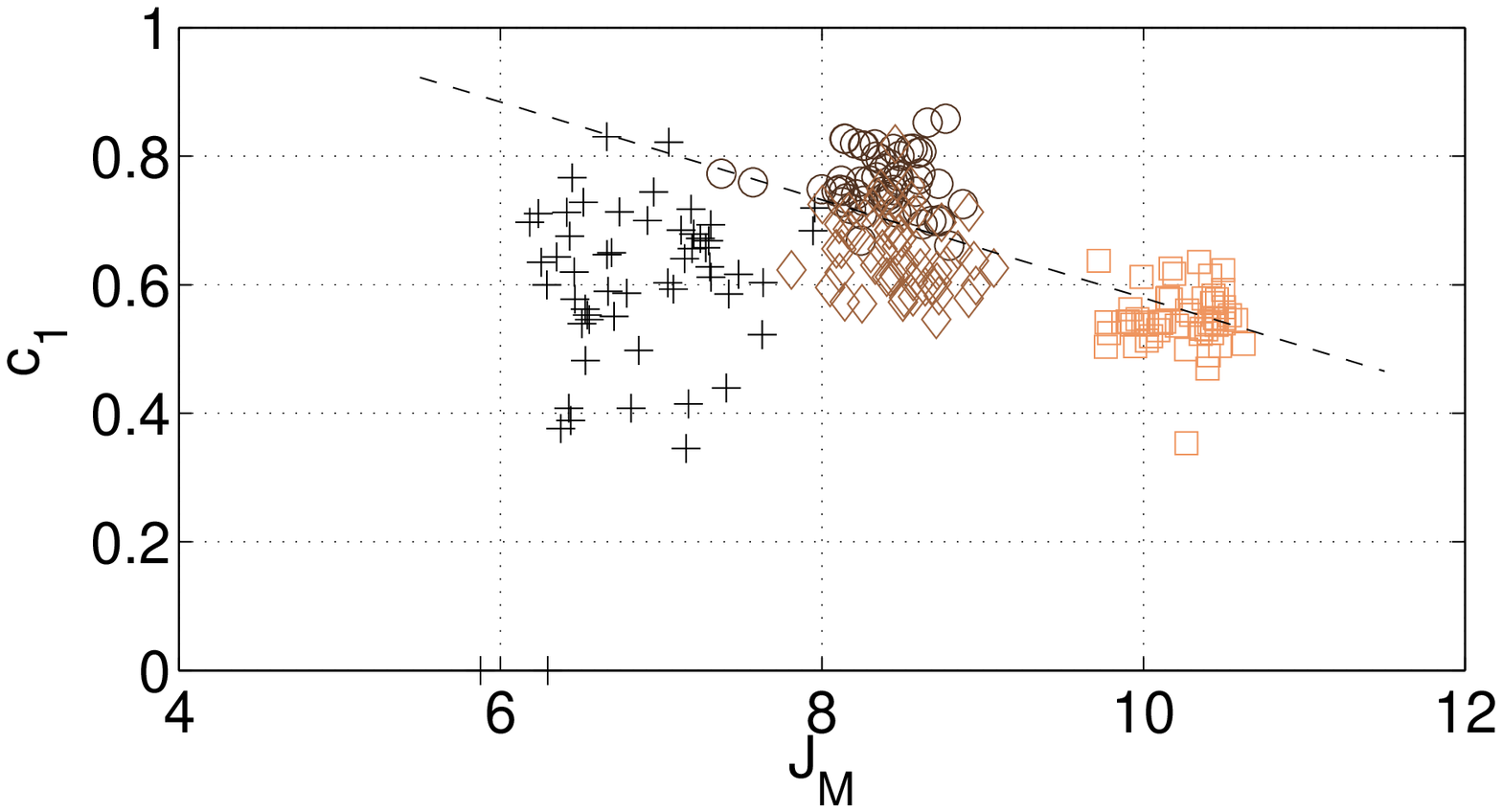}
                \label{fig:LRDRTTb}
                \end{minipage}
        }
        \,
        \subfloat[Top, Direct (upper right triangle) and partial (lower left triangle) correlations between RTT distribution and scaling parameters.  Bottom, corresponding Graphical Gaussian Model analysis of partial correlations.]{
         \begin{minipage}{0.3\textwidth}
	  \centering
	   \resizebox{\textwidth}{!}{
	    \begin{tabular}{|l|r|r|r|r|r|r|}
	    \hline
				& $J_R$ & $J_{M}$  & $J_F$ &  $H$  &  $c_1$ &  $c_2$ \\ \hline
	    $J_R$           &  -       & 0.92           & 0.49  & -0.07 & -0.07  &  0         \\ \hline
	    $J_{M}$   & -0.90 & -                &  0.57  &  -0.10 &  -0.13  &  -0.05 \\ \hline
	    $J_F$            &  0     &  -0.36              & -         &   0.15 & -0.07   &   0   \\ \hline
	    $H$             & 0      & 0                &  -0.25       & -         &  0.13 &  0.08 \\ \hline
	    $c_1$          & 0      & 0.18           	   & 0              & 0       & -       & -0.21         \\ \hline
	    $c_2$         &  0 &   0.19                  &  0            &  0       & 0.25 & -            \\ \hline
	    \end{tabular}
	   }
\vskip8mm
	    \includegraphics[width=\textwidth]{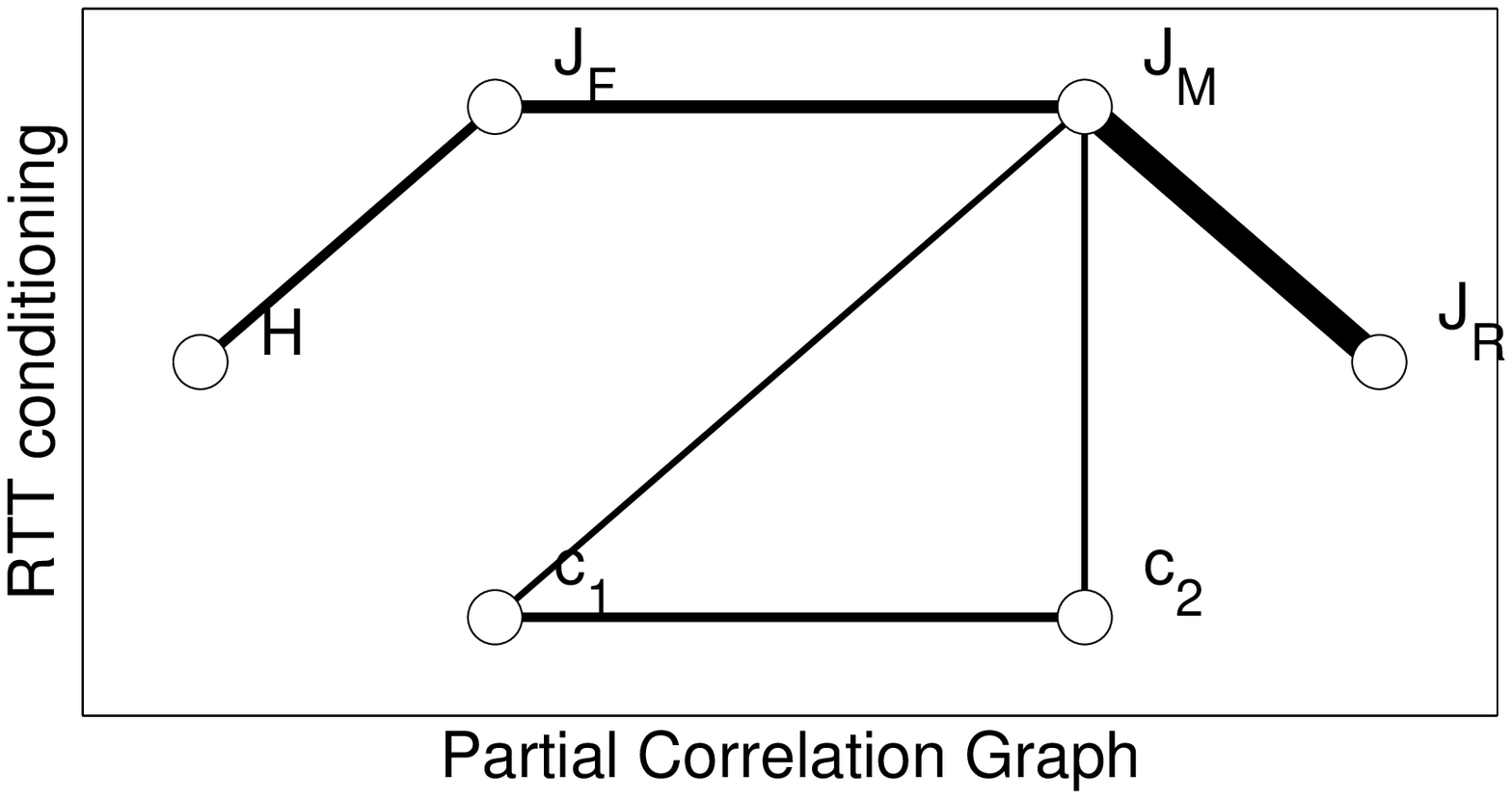}
	    \vskip2mm
	    \label{fig:LRDRTTc}
	  \end{minipage}
        }
         \vskip-8mm
        \caption{Origins of scaling at FS: RTT}\label{fig:Halpha}
\end{figure*}

\subsection{Fine scales}
\label{sec:FS}

\subsubsection{Nature of Scaling}

The parameter $H$ measured at FS should theoretically not be referred to as the Hurst parameter, which is in essence associated to LRD, a CS property. 
Yet, $H$, measured as a scaling exponent across FS, preserves the key interpretation of accounting for a scaling property of the correlation, though within a finite range of (fine) scales.
It is thus from now on labeled $ h $, 
in reference to its link to
the H\"older exponent, to which it should rather be associated in a multifractal setting (cf. e.g., \cite{WENDT:2007:E}). 

Fig.~\ref{fig:timeevol3D} (bottom row) shows that, for the 6-hour blocks of the 3-day trace, $h \simeq 0.70 \pm 0.02$ and that $c_2 \simeq -0.025 \pm 0.013$, with clear departure from $c_2 = 0$. 
Fig.~\ref{fig:timeevol15min} (bottom row) shows that FS scaling parameters on the 15-min trace along the 14 years take roughly piecewise constant values
 in the two periods separated by the link upgrade mid-2006: $h \simeq 0.57 \pm 0.03 $, $ c_2 \simeq -0.017 \pm 0.012 $ before 2006, and 
$h \simeq 0.64 \pm 0.03 $, $ c_2 \simeq -0.044  \pm 0.019 $ after 2006.
For both periods, $ H \simeq c_1 + c_2 $ and the latter estimates are satisfactorily consistent with those measured on the 3-day trace collected in 2013.

Global-LDs $c_2$ are found close to $0$ (both for the 6-hour blocks of the 3-day trace and consistently 
for the 14 years of 15-min traces), which, if taken for granted, would lead to conclude that scaling at FS traffic is not multifractal, 
However, Median-LD  $c_2 $ are consistently strictly negative along the 14 years and for the 3-day trace, with confidence intervals either obtained by bootstrap or as an average across sketch outputs, excluding $0$.
This unambiguously suggests that Internet scaling at FS is better described by multifractal models, than by LRD ones.
This is consistent with a number of contributions reporting multifractality in Internet traffic at scales of the order of 100ms on numerous different types of traffic and networks \cite{TTW97,Walter_MF_Allerton,Feldmann1998,Feldmann1998b,MasugiTakuma2007,LOISEAU:2010:A,Jaffard2015}.
Conversely, Ref. \cite{pseudoMF} reported a lack of evidence for multifractality in Internet traffic. 
However, in all earlier studies (including ours \cite{pseudoMF}), multifractality was analysed without recourse to random projections that brings robustness, and with tools (such as wavelet coefficients or increments) 
that are now known to have a low ability to discriminate $c_2 < 0$ from $c_2 = 0$ and thus are not good to unambiguously assess multifractality (see a contrario \cite{Jaffard2015}).

\subsubsection{Origins}

Temporal burstiness of Internet traffic time series has been consistently reported, with bursts occurring over many different scales 
ranging from tens to hundreds of milliseconds (cf.\,e.g., \cite{JD2005,lbfe2009}).  
Multifractality naturally provides a relevant framework to model temporal burstiness  (e.g., \cite{riedi}). 
Fig.~\ref{fig:timeevol15min} (bottom right plot) further shows that the link upgrade in 2006 does not create nor obliterate multifractality at FS. 
Yet, the link update, and corresponding increase of bandwidth, induce a change in scaling parameters that, though not large in amplitude, appears as clear and robust: 
Stronger global structure in the correlation of Internet traffic at FS (increase of $h$ and $c_1$) with yet larger variability beyond correlation, i.e., increased burstiness (increase of $  |c_2| $).
Multifractal scaling over FS constitutes an important statistical feature of Internet traffic, with notable impact of network performance (cf. e.g., \cite{cascadeperf,Feldmann1999}).
However, in contradistinction to CS, no clear and well recognized mechanism has been proposed to explain scaling at FS. 

While scaling at CS has been related to flow sizes, scaling at FS has rather been associated to packet injection mechanisms, that is essentially to the
TCP congestion control, designed to regulate traffic. 
For instance, \cite{JD2005} described how TCP self-clocking shapes the packet interarrivals within TCP connections and thus the FS temporal dynamics. 
TCP has thus been envisaged as one of the mechanisms potentially producing or modifying burstiness and hence multifractality or scaling at FS  \cite{Feldmann1998b,ERVW2002,Feldmann1999,Zhang2003,hohn03}.
The importance of time scales below the Round-Trip-Time (RTT) in Internet traffic temporal dynamics, has been evidenced  (e.g. \cite{JD2005}). 
The relation and strength of scaling with respect to other queuing mechanisms (such as bottlenecks and congestions) has been further documented \cite{Veres2000}.
In \cite{Feldmann1999}, it was shown that protocol related burstiness contributed strongly to the form of the LDs over (what corresponds here to) FS, as did network topology aspects, albeit in a very simple dumbbell topology.
In \cite{VKMV2003}, it was shown that TCP congestion control can propagate scaling between distant areas of the Internet. 
In \cite{LOISEAU:2010:A}, varying TCP parameters was shown empirically to modify
the scaling parameters measured at FS in Grid Traffic.
TCP essentially relies on modifying a time window injection mechanism, depending on bandwidth availability and traffic congestion status: 
Large available bandwidth yields an additive increase of the slow-start window per returning acknowledgment packet, 
while loss detection results in a multiplicative decrease. 
This mechanism, linked to a cascade mechanism, has been envisaged in the context of Internet traffic in \cite{riedi,VKMV2003} as a potential explanation for multifractality, together with the protocol hierarchy of IP data networks \cite{Feldmann1998b,Feldmann1999}.
To quantitatively investigate the relations between the RTT induced time scale and multifractality at FS, RTT has been estimated for each flow using Karn's algorithm \cite{karn:sigcomm87}, see also \cite{fontugne:infocom2015} for further details.
Here, the difference between each TCP packet transmission time and the corresponding acknowledgment reception time is measured.
Retransmitted packets are ignored to avoid ambiguous acknowledgments. 
The typical flow-RTT is estimated as the median of such RTTs.
For one example 15-min trace, the empirical distribution of RTT estimates is reported in Fig.~\ref{fig:LRDRTTa} 
and appears to be widely spread with several modes.  
Median-LDs are computed from subtraces designed by conditioning on RTT (partitioned in 4 classes as shown in Fig.~\ref{fig:LRDRTTa}).
They clearly suggest  that the frontier scale $\Delta_F = \Delta_0 2^{J_F}$ increases with RTT $\Delta_R = \Delta_0 2^{J_R}$. 
Fig.~\ref{fig:LRDRTTa} also  illustrates that the lower limit of the FS range is controlled by the packet IAT. 

This RTT-conditioning procedure is applied to 100 randomly chosen 15-min traces.
Parameters $H, h, c_1, c_2, J_F$ are estimated from the Median-LDs computed across the resulting $ 4 \times 100$ traces.
For each of the  classes, the median of $J_R$ is also measured and its dispersion, $J_{M}$, 
is estimated (as the median absolute deviation).
As expected theoretically, $c_1$ and $ h$ are highly correlated ($\rho = 0.78$) as they are essentially measuring the same dynamical property at the 2nd-order statistic level.
The latter is thus removed from further analyses for clarity of exposition. 
The Graphical Gaussian Model framework \cite{Wang2009_bayesian} is used to assess direct and partial correlations amongst the remaining 6 parameters.
Partial correlation is classically used to quantify how much dependency remains between two variables, once indirect correlations induced by the other variables are removed.
This leads to the graph of relations reported in Fig.~\ref{fig:LRDRTTc} and to the following comments and conjectures.  

i) A weighted least square regression of $ J_F $ against $J_R$ (reported in Fig.\ref{fig:LRDRTTb}) indicates that they are significantly correlated and that $J_F \simeq J_R $ (or $\Delta_F \simeq \Delta_R$), in clear agreement with daily-median RTT reported in Fig.~\ref{fig:FS} for the 1176 traces. 
Further, $J_F$ shows significant partial correlations with both $H$ and $J_M$.
RTT can be interpreted as a specific scale of time, characteristic of a flow, and resulting jointly from interactions between available bandwidth, flow size and destination address. 
This typical time scale thus breaks the CS scaling induced by heavy tails thus creating the frontier scale $J_F$, whose value hence
results from a competition between the CS heavy tail and the FS packet injection mechanisms. 

ii) The median RTTs $J_R$ is strongly correlated to its dispersion $J_{M}$ but shows negligible partial correlations with any other scaling parameter. 
Conversely, the intra-class RTT dispersion $J_M$ shows significant partial correlations with all scaling parameters. 
The large dispersion of RTTs (even within classes) implies that a broad spread of time scales contributes to temporal dynamics.
A breadth of time scales, with no distinguishable roles, contributing to temporal dynamics constitutes one potential known generic mechanism inducing scaling. 
Further, partial correlations suggest that $J_M$ acts as a hub controlling the values of $c_1$ (thus $h$) and $c_2$.
Negative direct correlations (Figs.~\ref{fig:LRDRTTb} and \ref{fig:LRDRTTc}) between $c_1$ or $c_2$ and both $J_R$ and $ J_M $ indicate that a decrease of $J_R$ and $ J_M $ induces larger $c_1$ and $|c_2|$.
This confirms that the modulation of both $c_1$ and $c_2$ also stems from the competing CS/FS mechanisms.
 A decrease in $J_R$ and $ J_M $ may be a consequence of an increase in available bandwidth. 
For instance, the link upgrade in 2006, resulting into an increase of the available bandwidth, also implies a decrease in $J_F$ and $J_R$.
This bandwidth increase makes possible a much more active and efficient TCP control mechanism, thus larger temporal burstiness  and richer FS temporal dynamics, which are quantified by an increase of both $c_1$ (stronger correlation) and $|c_2|$ (stronger multifractality).

iii) $H$ at CS can be interpreted as not impacted by packet injection mechanisms and thus as solely depending on flow heavy tails. 

\subsubsection{Partial conclusion 5}

The present investigations provide robust  
evidence that the multifractal paradigm offers a relevant description of Internet traffic temporal dynamics at FS, notably accounting for burstiness, consistently along the 14 years studied here, as well as along the 3-day trace.
They also report quantitative and consistent empirical evidence, relating the frontier scale $\Delta_F $ and FS temporal dynamics to RTT $\Delta_R $ distributions, and thus to the TCP mechanism, in competition with the heavy tail CS mechanism. 
\vspace{-2mm}

\section{Discussions and conclusions}  
\label{sec:conc}

This longitudinal study over 14 years and across 3-days suggests the following comments and conclusions related to the scaling properties of Internet (MAWI) traffic, and echoing the challenges raised in Introduction section.

A relevant study of scale invariance in Internet traffic cannot be achieved without the combined use of multiscale representations and random projections. 
The latter permit statistical analyses that are robust to anomalous behaviors and thus describe background regular traffic. 
The former allow the range of scales where scaling behavior holds (through wavelet coefficients) to be determined and allows for a better discrimination of the nature of the scaling (wavelet leaders enable one to discriminate multifractality, beyond LRD). 

Scaling properties in Internet traffic do exist and are not caused by spurious non stationarity. 
They develop not within a single scaling range but across two scaling ranges, the coarse and fine scales.
This biscaling regime is a robust property that holds within traffic monitored continuously along the 3-day trace. 
This biscaling regime thus provides practitioners with a paradigm to describe the statistical properties of Internet traffic over scales from ms ($ 10^{-3} $s) to several hours ($ 10^{4} $s), impressively ranging over 7 decades. 
This biscaling regime is also a property that has remained remarkably stable both in terms of the qualitative shape of LDs and in the values of the scaling exponents across the 14 years of the present study, thus showing that the amazing evolution of services, applications, behaviors and of technological capacity increase (bandwidth, volumes, \ldots) have not caused any significant changes in the temporal dynamics of Internet traffic time series. 
Particularly, in contradistinction to what has sometimes been proposed (cf. e.g., \cite{TerdikGyires,TerdikGyires2009}), the present study shows that neither the increase in bandwidth and traffic volume nor the constant evolution of applications and services,  have led to the disappearance of scaling properties.
They favor neither the disappearance of LRD at CS, nor the return to Gaussian statistics and reduced burstiness at FS -- multifractal properties remains.

The scaling at CS (above $1$s) is well described by LRD that models temporal dynamics at the covariance (2nd order statistics) level, with no need for recourse to multifractality. 
As often proposed qualitatively, we showed here an exceptional quantitative agreement with the heavy tail behavior of the flow size (number of packets) distribution, providing an empirical validation of the universal mechanism proposed in \cite{onoffproof}. 
This CS scaling holds up to the coarsest scale practically available for the analysis (several hours in the case of the 3-day trace) and remains visible down to scales of the orders of $0.5$s where competing mechanisms related to within-flow packet injection become dominant. 
Details of the CS/FS competition, governing in particular the definition of the corresponding scale ranges, depend on RTT.  
This forms a link between scaling properties and protocol-specific mechanisms which can be further explored in future work.

In the FS regime, from several hundreds of ms down to the typical packet IAT (below 1ms), scaling temporal dynamics are better described by multifractal properties. 
Notably, while it is sometimes incorrectly associated to LRD, the burstiness of Internet traffic time series is actually well accounted for by multifractality: the larger $|c_2|$ (multifractality) the more prominent the temporal burstiness. 
In contradistinction to CS, there is no universal mechanism proposed to described scaling at FS.
Packet injection policies are driven by several protocols, the most prominent of which being TCP.
TCP relies on a specific time scale, the RTT of a flow, that depends on flow size and destination, traffic volume and bandwidth.
This study showed that the 
RTT distribution is very broad (see also \cite{fontugne:infocom2015}), thus producing a large continuum of time scales contributing to temporal dynamics and hence scaling.
This study has provided quantitative evidence of the relations of scaling at FS with RTT. 
This is hence a possible mechanism producing scaling at FS, competing with scaling at CS, and thus producing the biscaling regime. 
These links to RTT and TCP do not per se explain multifractality. 
Therefore, the point made in the present contribution is not that Internet traffic is multifractal at FS, but rather that multifractal processes constitute an efficient modeling of the statistics of Internet traffic at FS, notably accounting for temporal burstiness. 

The frontier scale separating the two scaling regimes ranges from several hundreds of ms to $1$s.
It can be regarded as a typical time scale separating Internet users (human beings) generating/producing the contents transferred through the Internet and hence to some large extent, the heavy tail of flow size, and technological behaviors (packet injection protocols).   

Instead of having recourse to a different description for each scaling range, with no relation between the two ranges, one may prefer to use a single model valid across all scales (cf. e.g., \cite{lbfe2009} for the definition of an index of variability across all scales). 
The Cluster Point Process model (CPP), put forward in \cite{hohn03}, provides a unique description of traffic across all available scales in a hierarchical manner (clusters of point processes) that accounts for flows and packets
with-in flows. 
By construction, the CPP model is asymptotically LRD at CS when cluster size is
heavy tailed, but being a point process cannot be strictly multifractal in the limit of FS. 
However, the extent to which it is well approximated as such across a large range of FS is currently being investigated. 
While theoretically appealing to describe Internet traffic, the CPP model is fully parametric hence less versatile to accommodate  real data.
Multifractal, and scaling exponents $c_1$  and $c_2$, can thus be envisaged as alternative versatile semi-parametric features, practically, relevant and useful for various network tasks (e.g., traffic characterization and anomaly detection, cf. e.g., \cite{fontugne2015icassp}). 


{
\bibliographystyle{ieeetran}
\bibliography{MFTraffic_V2}
}
\begin{IEEEbiography}{Romain Fontugne} completed a Computer Science Ph.D. from Sokendai at the National Institute of Informatics (NII) in Tokyo (2011). He was a JSPS Postdoctoral Research Fellow at the University of Tokyo (2011-2013) and worked at NII for the NECOMA Project (2013-2015). 
Since 2015, he is a Senior Researcher at Internet Initiative Japan Inc. He is also a part time lecturer at Waseda University, and an active member of the WIDE Project and the Japanese-French Laboratory for Informatics (CNRS).
His current research interests include Internet measurements, traffic analysis, and Internet routing.
\end{IEEEbiography}
\vspace*{-1cm}
\begin{IEEEbiography}{Patrice Abry} completed a Ph.D. in Physics and Signal Processing, at Universit\'e Claude-Bernard University in Lyon in 1994. 
He is currently CNRS Senior Scientist, at the Physics dept. of Ecole Normale Sup\'erieure de Lyon.  
He received the AFCET-MESR-CNRS prize for best Ph.D. in Signal Processing for the years 93-94. 
He has been elected IEEE fellow in 2011 and serve for the IEEE SPS Signal Processing Theory and Methods Committee.
His current research interests include wavelet-based analysis and modeling of statistical scale-free and multifractal dynamics. 
Beyond theoretical developments, Patrice Abry shows a strong interest for real-world applications (hydrodynamic turbulence, computer network teletraffic, Heart Rate Variability, neurosciences or art investigations).
\end{IEEEbiography}
\vspace*{-1cm}
\begin{IEEEbiography}{Kensuke Fukuda} is an associate professor at the National Institute of Informatics (NII).
He earned his Ph.D degree in computer science from Keio University in 1999.
He worked in NTT laboratories from 1999 to 2005, and joined NII in 2006.
He was a visiting scholar at Boston University in 2002 and a visiting scholar at the University of Southern California / Information Sciences Institute in 2014-2015.
He was also a researcher of PRESTO JST (Sakigake) in 2008-2012.
His current research interests are Internet traffic measurement and analysis, intelligent network control architectures, and scientific aspects of networks.
\end{IEEEbiography}
\vspace*{-1cm}
\begin{IEEEbiography}{Darryl Veitch}
completed a BSc.~Hons.~at Monash University, Australia (1985)
and a mathematics Ph.D.~from DAMPT, Cambridge (1990). He worked at
TRL (Telstra, Melbourne), CNET (France Telecom, Paris), KTH
(Stockholm), INRIA (Sophia Antipolis and Paris, France), Bellcore (New Jersey),
RMIT (Melbourne), Technicolor (Paris) and EMUlab and CUBIN at The University of
Melbourne, where he was a Professorial Research Fellow until end 2014.
He is now a Professor in the School of Computing and Communications at the 
University of Technology Sydney. 
His research interests are centered around computer networking and inference and include traffic
modelling, parameter estimation, the theory and practice of active measurement, 
traffic sampling and sketching, information theoretic security, and clock synchronisation over networks.  
He is a Fellow of the IEEE.
\end{IEEEbiography}
\vspace*{-1cm}
\begin{IEEEbiography}{Kenjiro Cho}
    is Research Director at Internet Initiative Japan, Inc.
He is also an adjunct professor at Keio University, and a board member
of the WIDE project.
He received the B.S. degree in electronic engineering from Kobe University,
the M.Eng. degree in computer science from Cornell University, and the Ph.D.
degree in media and governance from Keio University.
His current research interests include Internet data analysis,
networking support in operating systems, and cloud networking.
\end{IEEEbiography}
\vspace*{-1cm}
\begin{IEEEbiography}{Pierre Borgnat}
is a Senior Researcher in Signal Processing at CNRS (Directeur de Recherche), and works at the Laboratoire de Physique, ENS de Lyon, France, since 2004. He is Director of IXXI (Complex System Institute Rhône-Alpes) since the end of 2016 and was Deputy-Director in 2015-2016. He obtained the Agr\'egation in Sciences Physiques in 1997, a Ph.D. degree in physics and signal processing from ENS de Lyon in 2002, and a ``Habilitation \`a diriger des recherches'' from the ENS de Lyon in 2014. He is AE of the Trans. on Signal Processing since 2015. His research interests are in statistical signal processing, for nonstationary processes, scaling phenomena, for graph signal processing, and complex networks. He works on many applications, including Internet traffic modeling and measurements, fluid mechanics, analysis of social data, and transportation studies. 
\end{IEEEbiography}
\vspace*{-1cm}
\begin{IEEEbiography}{Herwig Wendt} received the Ph.D. degree in Physics and Signal processing from Ecole Normale Supérieure de Lyon in 2008. From October 2008 to December 2011, he was a Postdoctoral Research Associate with the Department of Mathematics and with the Geomathematical Imaging Group, Purdue University, West Lafayette, Indiana. Since 2012, he is tenured research scientist with the Centre National de Recherche Scientifique (CNRS) and with the Signal and Communications Group of the IRIT Laboratory, University of Toulouse.
His research interests include the statistical analysis and modeling of scale-free phenomena and multi-scale analysis and computation.
\end{IEEEbiography}

%

\end{document}